\documentclass{aa}
\usepackage{natbib,float}
\usepackage{graphicx,booktabs}
\usepackage{multirow}
\usepackage{color}
\usepackage{txfonts}

\newcommand{\co}{CO(2$-$1)}
\newcommand{\kms}{\mbox{km\,s\ensuremath{^{-1}}}}
\newcommand{\msun}{\emr{M_{\odot}}}
\newcommand{\farc}{\mbox{\ensuremath{^{\prime\prime}}}}
\newcommand{\rco}{CO(2$-$1)/CO(1$-$0)}
\newcommand{\emr}[1]{\ensuremath{\mathrm{#1}}}

\newcommand{\FigMomentsFull}{
  \begin{figure*}
    \centering{} %
    \includegraphics[width=0.9\linewidth]{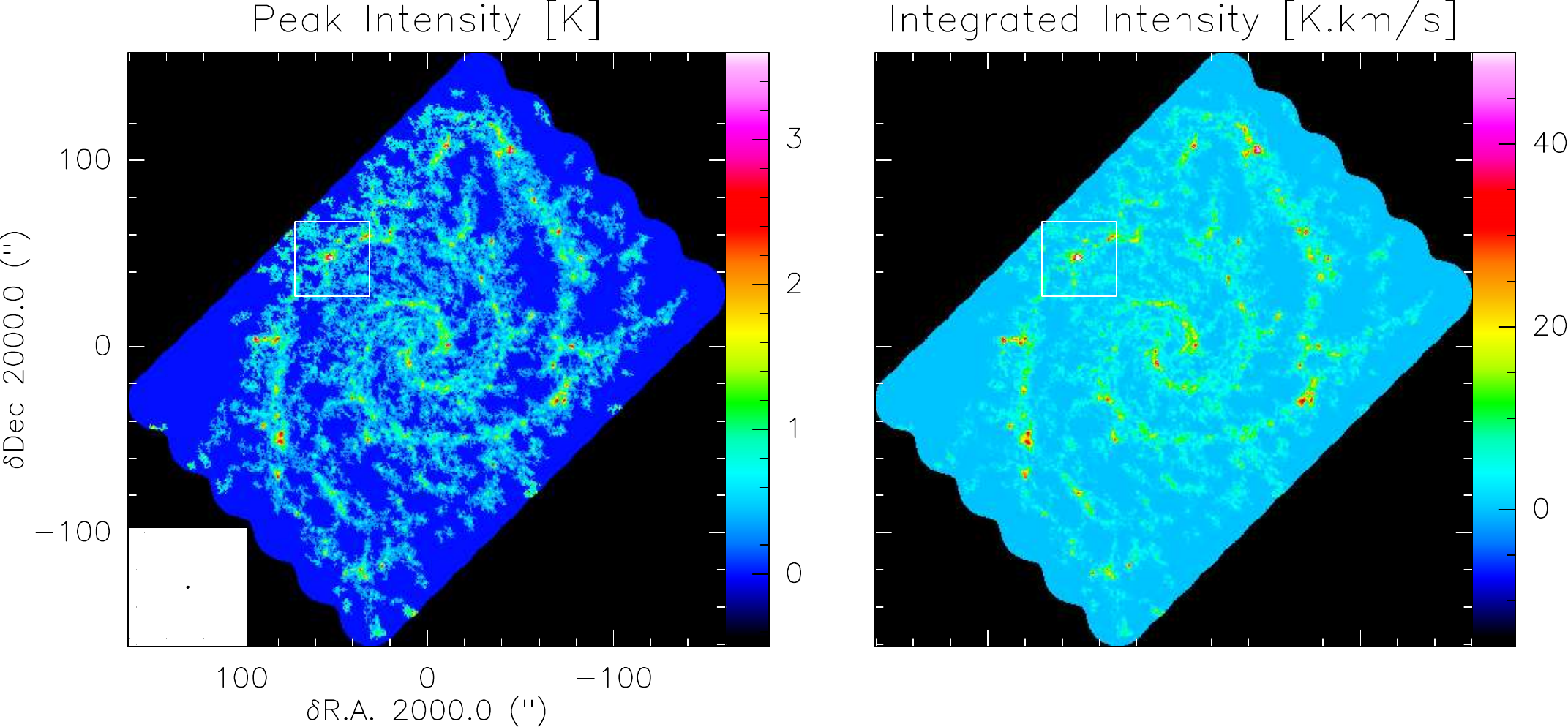}
    \includegraphics[width=0.7\linewidth]{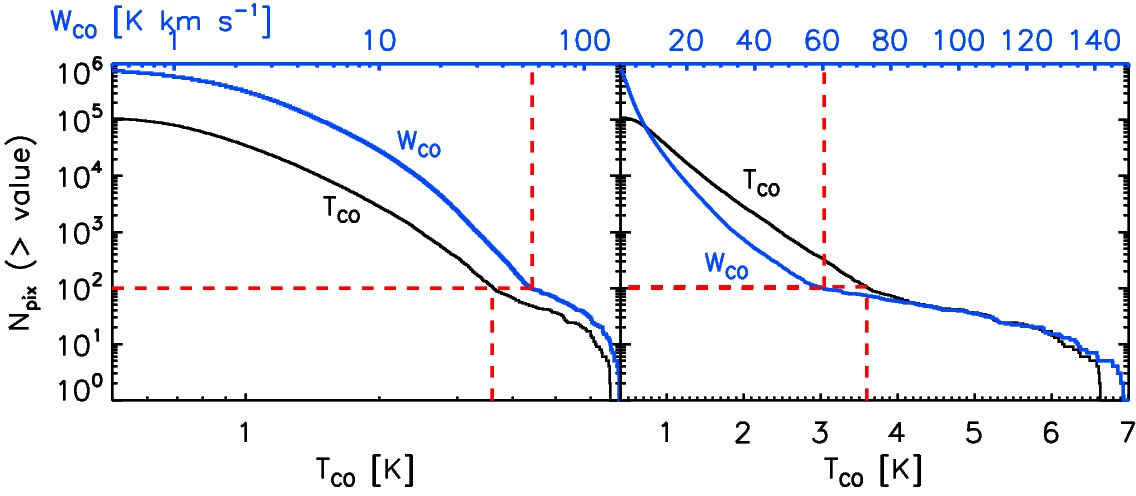}
    \caption{\textbf{Top:} Spatial distribution of the peak temperature
      (left) and integrated intensity (right) of the \co{} line over the
      full observed field of view of NGC\,628. The color scales a dynamic
      range of the peak temperature and integrated intensity, saturated to
      high intensity of 3.6~K and 45~K~\kms. This is a compromise
        to show the spatial distribution of the $\sim99.99\%$ of the pixels
        even though this may give the incorrect impression that other
        clouds in NGC\,628 are as bright as the headlight cloud. The white
      square shows the position of the headlight cloud in NGC\,628, which
      is zoomed in the left part of Fig.~\ref{fig:moments:zoom}. Offset
      positions are from the galactic center, $\alpha:$
      01$^h$36$^m$41$\fs$72, $\delta:$ +15$\degr$46\arcmin59$\farcs$3
      J2000. The beamsize is shown in the bottom left corner of the left
      panel. \textbf{Bottom:} Number of pixels whose \co\ peak temperature
      $T_{\rm CO}$ (black curve) and integrated \co\ intensity $W_{\rm CO}$
      (blue curve) are above the value given in the bottom abscissa for
      $T_{\rm CO}$ and in the top abscissa for $W_{\rm CO}$, for the entire
      galaxy. The left panel shows the full range of values in logarithmic
      scale while the right one zooms in linear scale. The red dashed lines
      show the values corresponding to 100 pixels (or a surface of about
      $2''\times 2''$) in the CO map.}
    \label{fig:moments:full}
  \end{figure*}}

\newcommand{\FigZoom}{
  \begin{figure*}
    \centering %
    \includegraphics[width=0.475\linewidth]{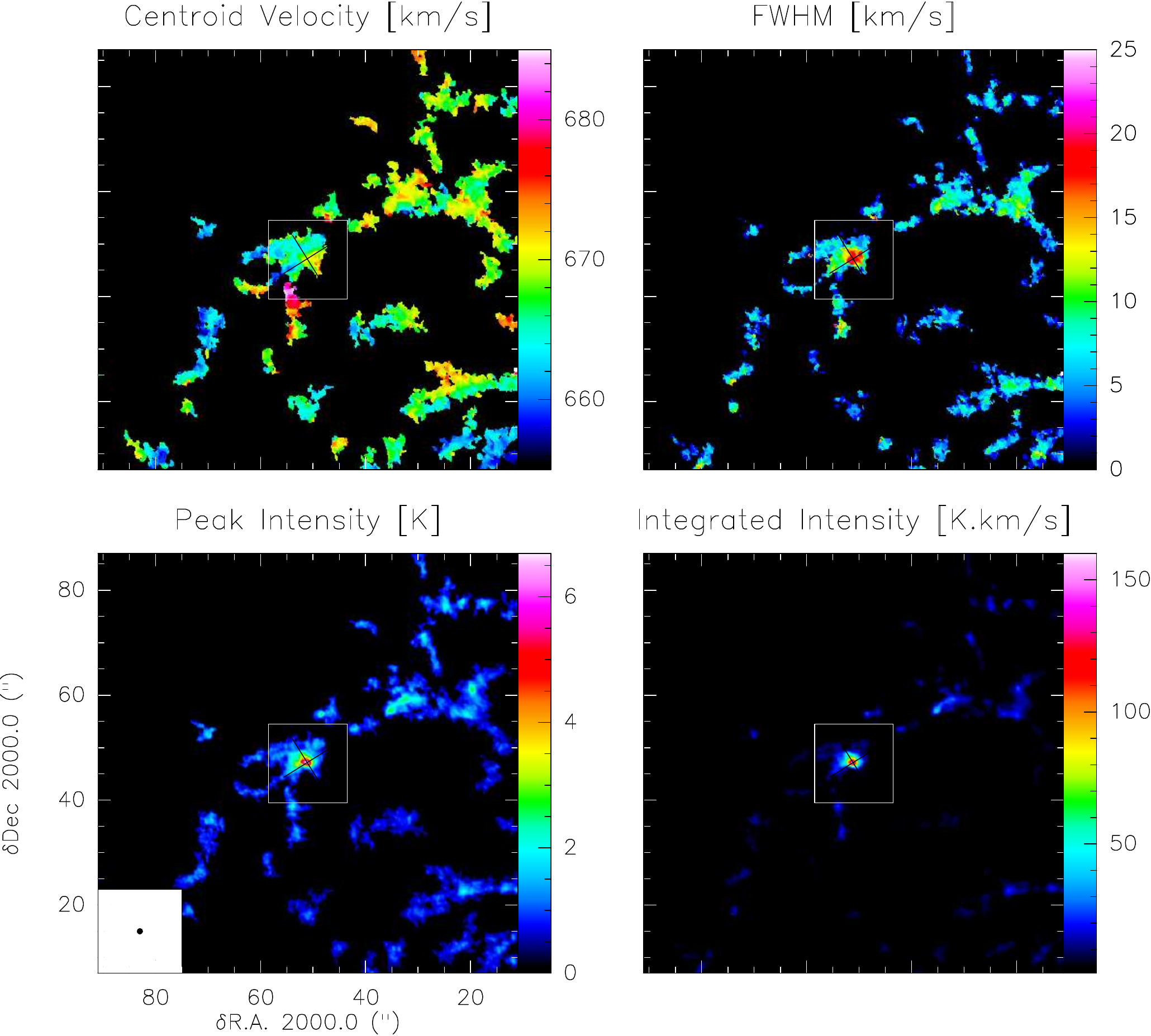}
    \hfill{}
    \includegraphics[width=0.475\linewidth]{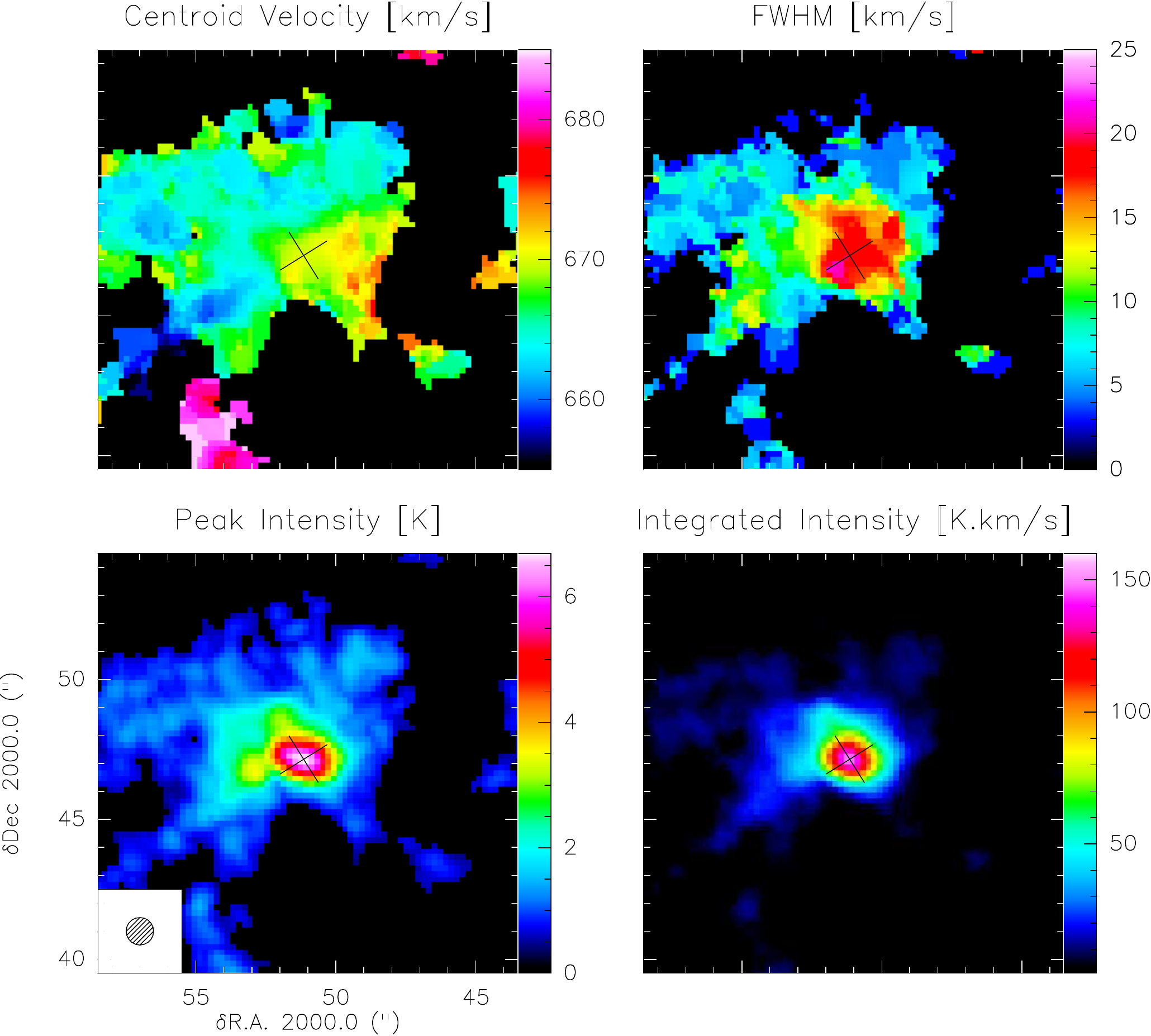}
    \caption{Spatial distributions of the moments of the \co{} line
      centered over the headlight cloud. The 4 left panels show how this
      molecular cloud relates to the spiral arm, covering a field-of-view
      of 80\arcsec$\times$80\arcsec, while the 4 right panels zoom in on
      the molecular cloud itself. The beamsize is shown in the bottom left
      corner of the bottom left panels.}
    \label{fig:moments:zoom}
  \end{figure*}}

\newcommand{\FigLarson}{
  \begin{figure*}
    \centering \includegraphics[width=0.72\linewidth,trim={0 0 0
      0},clip]{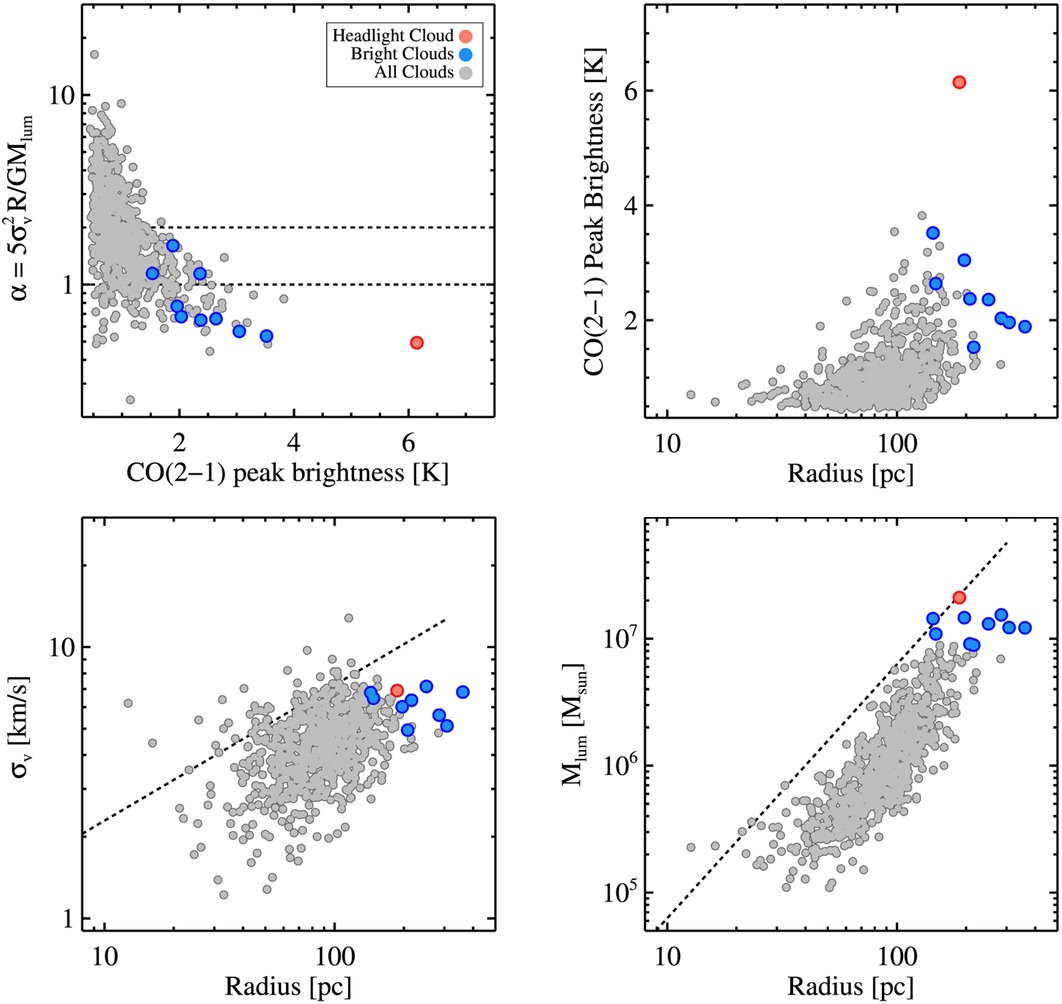}
    \caption{Scaling relations between physical properties of the molecular
      clouds in NGC\,628: a) virial parameter as a function of CO peak
      brightness, b) CO peak brightness as a function of cloud size, c)
      linewidth as a function of size, and d) luminous mass as a function
      of size. In all panels, the headlight cloud is the red point, and the
      nine next most luminous ($M_{\rm lum}>8.6~\times~10^6\,$\msun)
      molecular clouds are blue. The horizontal dashed lines in the top
      left panels represent virial parameters of $\alpha=1$ and $\alpha=2$,
      values that are indicative of virialized and self-gravitating clouds
      respectively. The dashed line in the bottom left panel represents the
      size-linewidth relation reported by \citet{solomon87} for inner Milky
      Way clouds.  The dashed line in the bottom right panel represents a
      mass surface density of
      $\Sigma_{\rm cld} = 200\,$M$_{\odot}$\,pc$^{-2}$.}
    \label{fig:larson}
  \end{figure*}}

\newcommand{\FigStacks}{
  \begin{figure*}
    \centering \includegraphics[width=0.45\linewidth,trim={0 0 0
      0},clip]{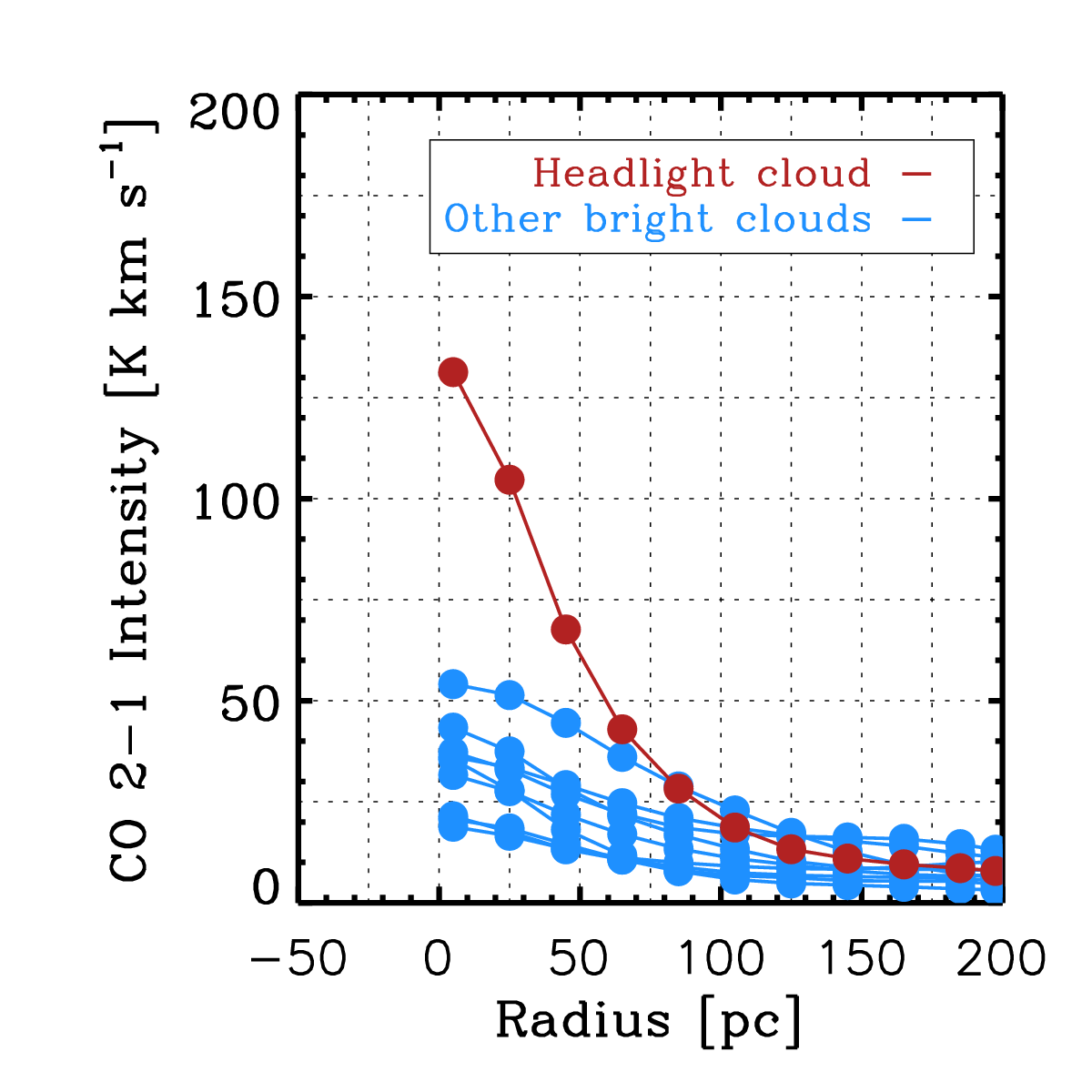}
    \includegraphics[width=0.45\linewidth,trim={0 0 0
      0},clip]{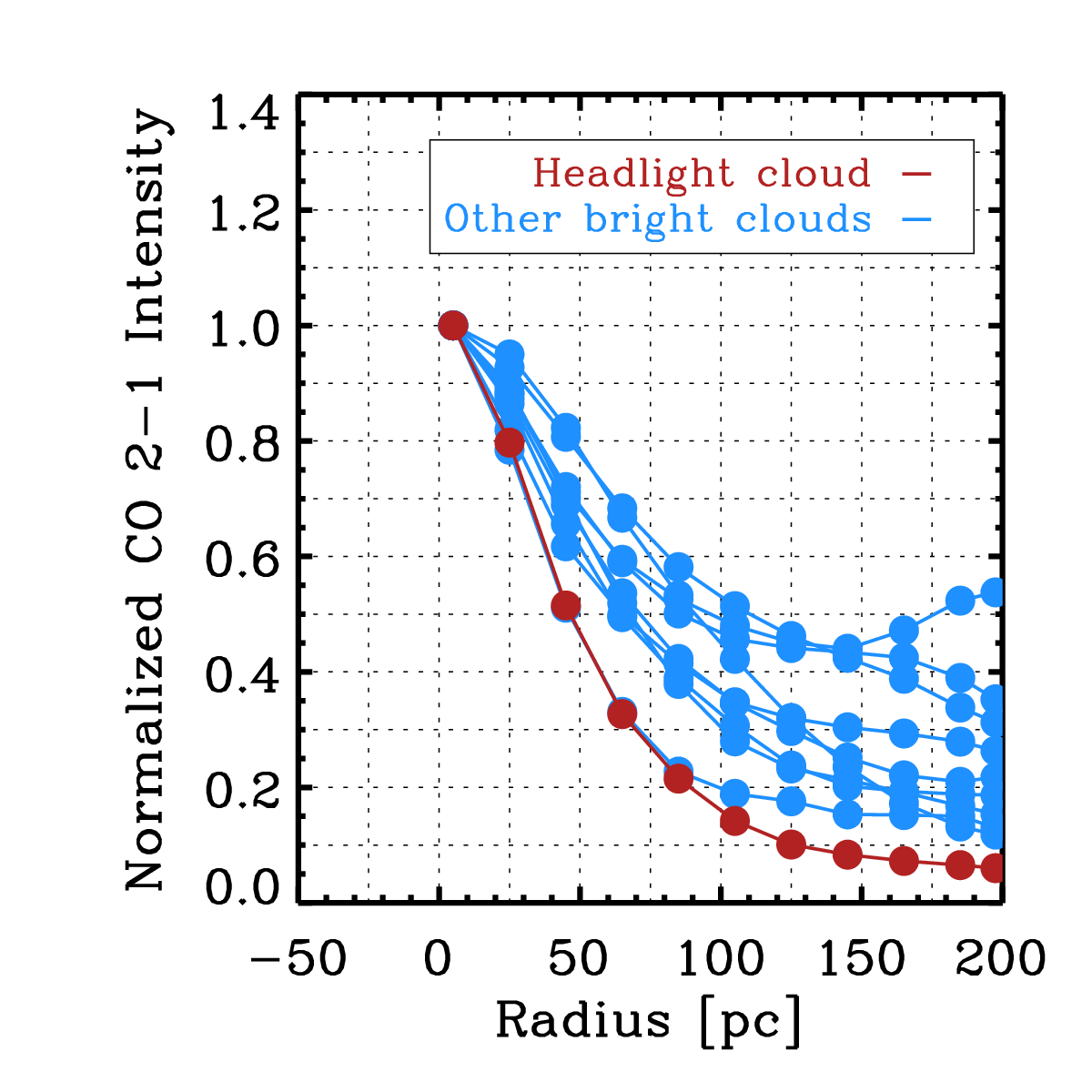}
    \includegraphics[width=0.45\linewidth,trim={0 0 0
      0},clip]{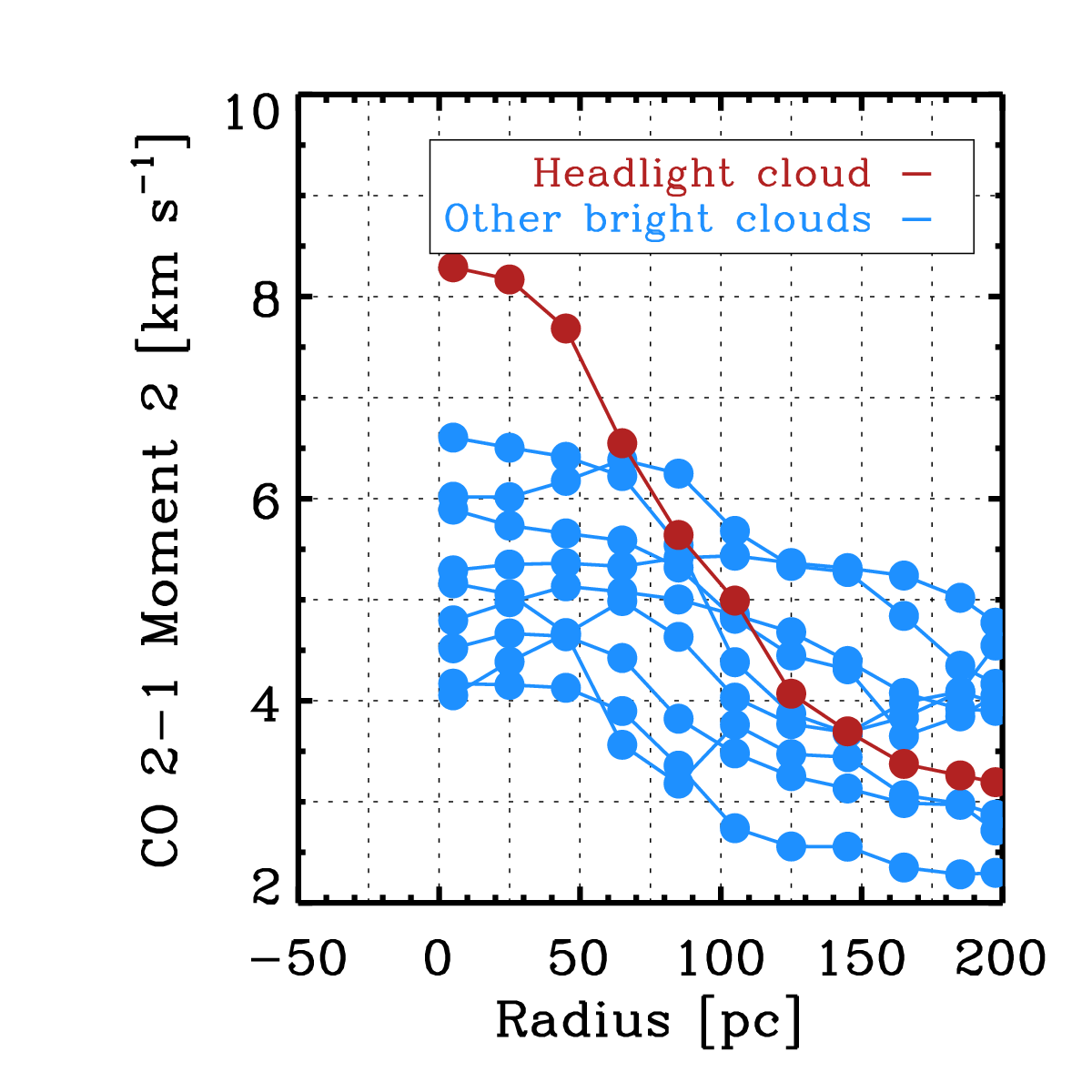}
    \includegraphics[width=0.45\linewidth,trim={0 0 0
      0},clip]{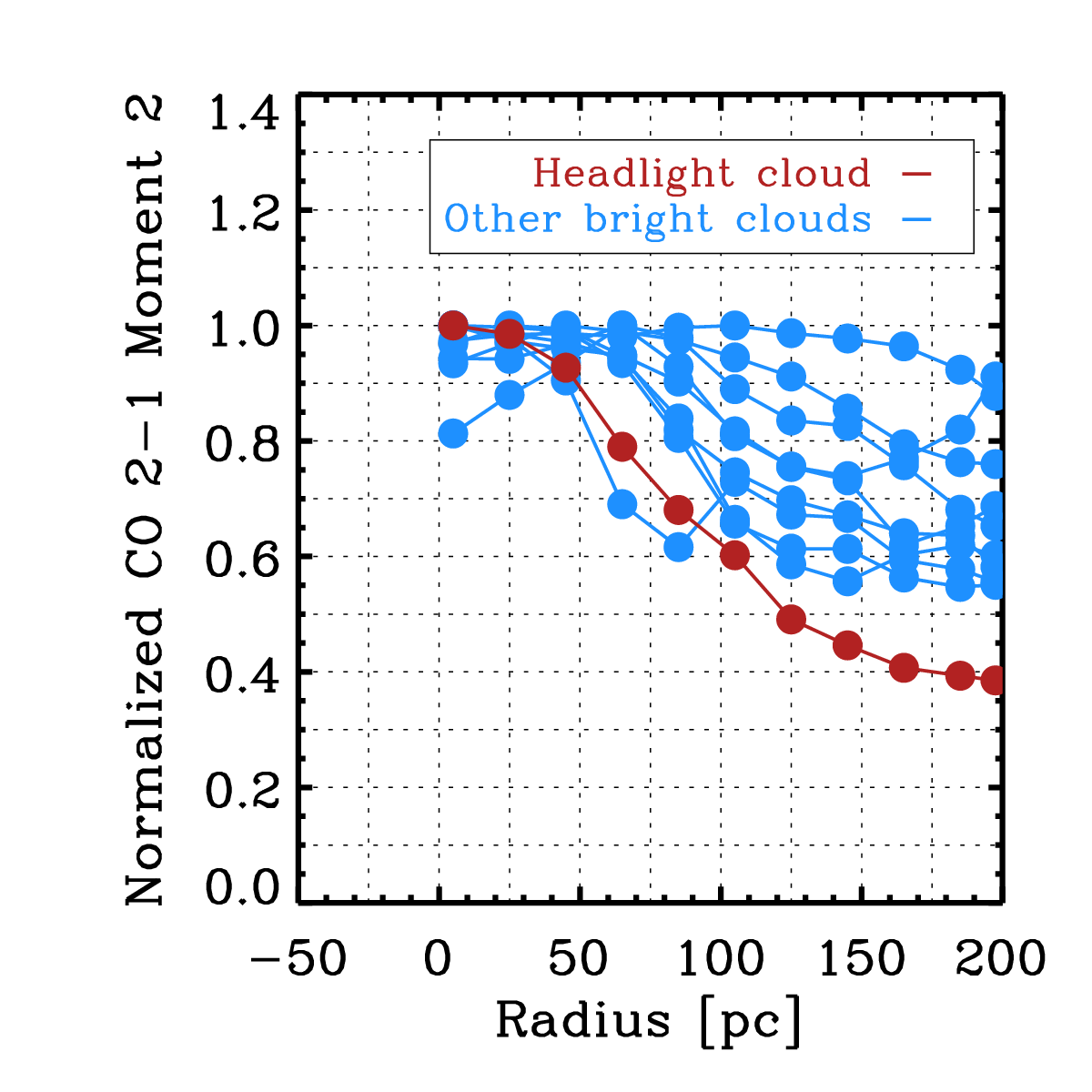}
    \caption{Profiles of mean integrated CO~(2$-$1) intensity (top row) and
      mean CO~(2$-$1) line width (bottom row) around the 10 most luminous
      clouds in NGC\,628, with the headlight cloud highlighted in red. We
      use the second moment to characterize the line width, but note that
      other possible line width diagnostics (e.g. equivalent width) show
      similar trends to those plotted in the bottom row. For each cloud, we
      measure the azimuthally averaged quantity in radial bins centered on
      the cloud peak. The core of the headlight cloud is more than twice as
      bright as the next brightest massive cloud, and also has an enhanced
      central line width (left panels). In the right panels, we highlight
      the relative shape of the radial profiles by normalizing all
      measurements by the value at the cloud's central position. The
      headlight cloud is more compact than the other massive clouds, though
      still moderately extended compared to the beam.}
    \label{fig:stacks}
  \end{figure*}}

\newcommand{\FigLinesLVG}{
  \begin{figure*}
    \centering %
    \includegraphics[width=0.45\linewidth]{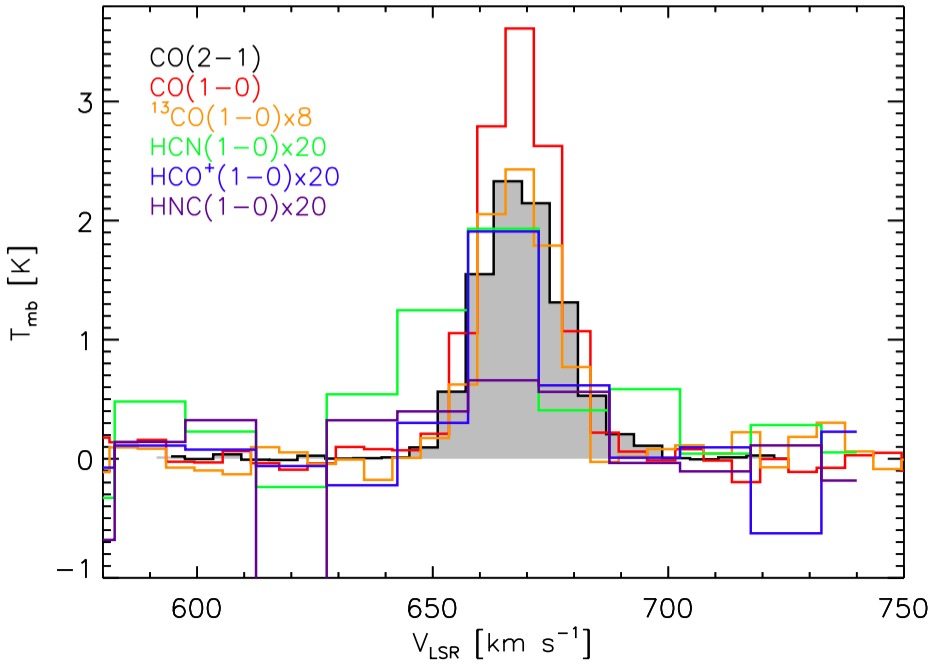}
    \includegraphics[width=0.45\linewidth]{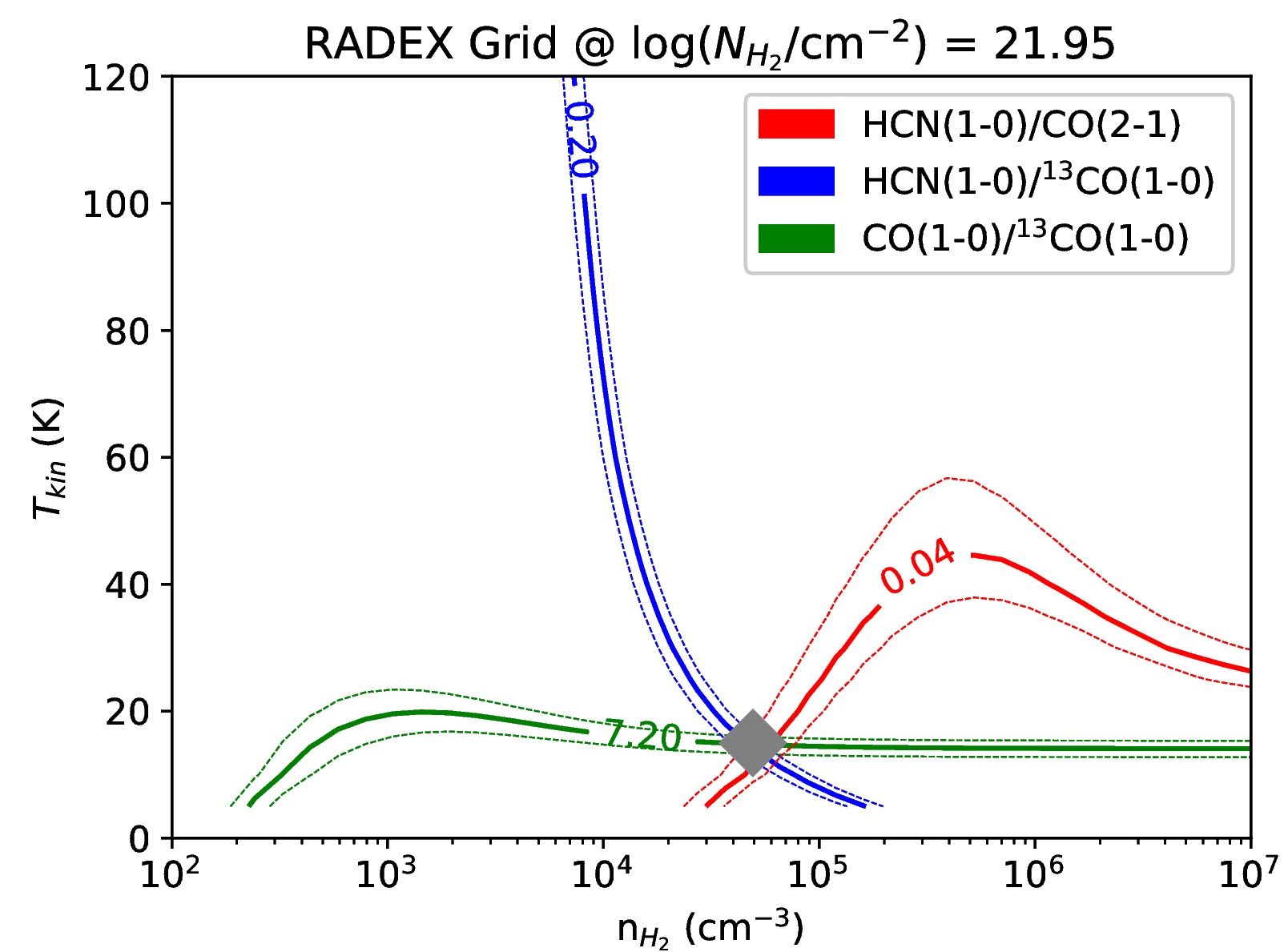}
    \caption{{\bf Left:} Line profiles of all measured molecular gas
      tracers extracted at the position of the peak emission of
        each line, after convolving all data to the angular resolution of
      the $^{13}$CO(1$-$0). The CO(2$-$1) spectrum appears as the gray
      filled region. Other lines are scaled by the factors indicated in the
      legend to put them on the same plot. {\bf Right:} RADEX output
      showing fits to the observed ratios among these lines. Blue, green,
      and red lines correspond to the observed HCN/$^{13}$CO,
      HCN(1$-$0)/CO(2$-$1) and $^{12}$CO/$^{13}$CO line ratios. The gray
      diamond marks the position on the RADEX [$n_{\rm H_2}$,
      $T_{\rm kin}$] grid where the $\chi^2$ is minimal, the best fit
      values. The dashed lines show the 95\% confidence level of the
      $\chi^2$ fit that is smaller than the size of the diamond.}
    \label{fig:lvg}
  \end{figure*}}

\newcommand{\FigLineRat}{
  \begin{figure}
    \centering %
    \includegraphics[width=\linewidth]{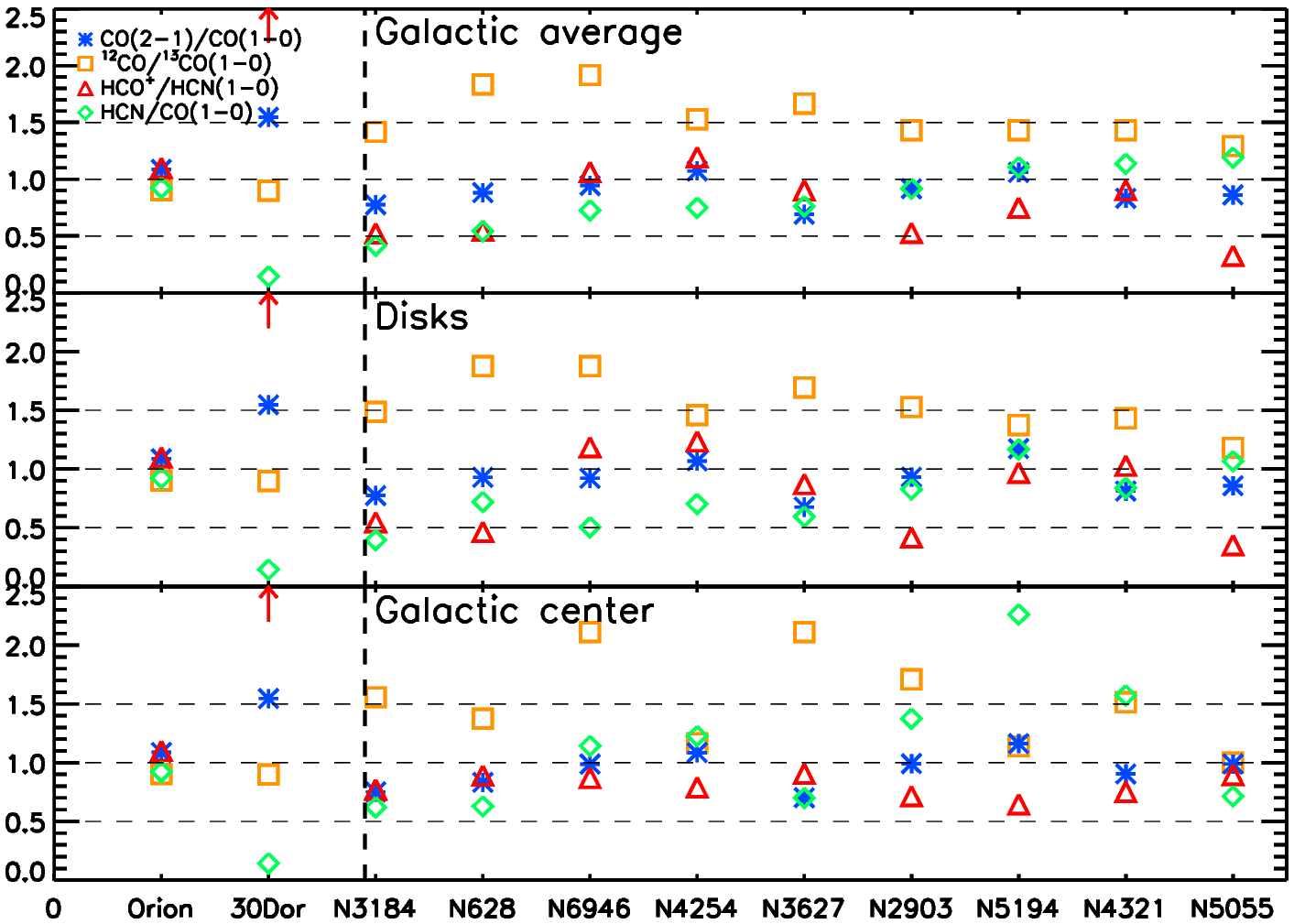}
    \caption{Line ratios observed in different sources, normalized to the
      values measured for the headlight cloud in NGC\,628. Yellow, blue,
      red and green symbols correspond to the $^{12}$CO/$^{13}$CO, \rco\,
      HCO$^+$/HCN and HCN/CO(1$-$0) line ratios. The values given for
      30~Dor correspond to the values in the 30~Dor-10 molecular cloud. The
      red arrow for 30~Dor marks that the HCO$^+$/HCN ratio is higher than
      2.5. See Table~\ref{tab:rat-gals} for the absolute values for these
      line ratios and their references.}
    \label{fig:rat-gals}
  \end{figure}}
    
\newcommand{\FigRCOMass}{
  \begin{figure}
    \centering %
    \includegraphics[width=0.8\linewidth,clip]{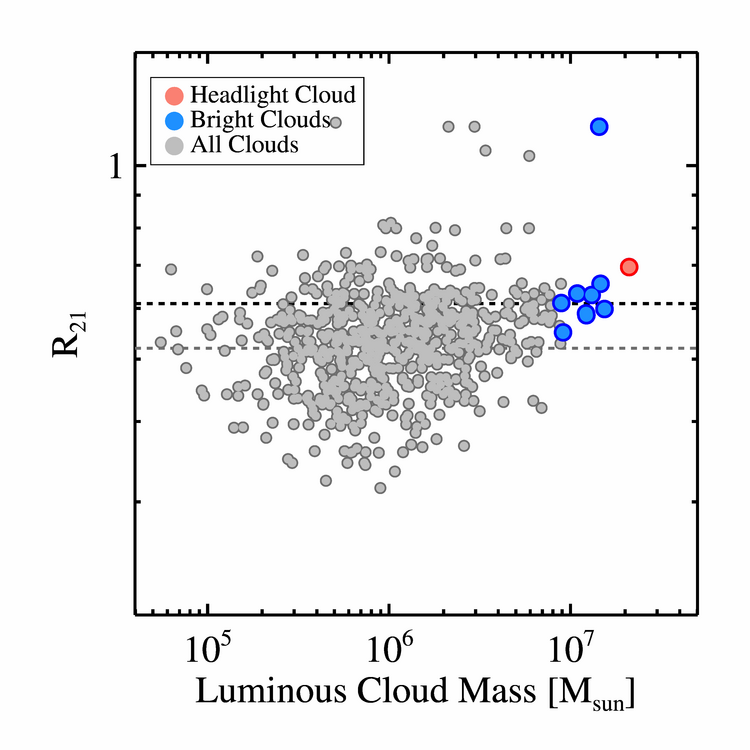}
    \caption{The \rco\ ratio measured on 140\,pc scale at the positions of
      clouds in NGC\,628. As in Figure~\ref{fig:larson}, the headlight
      cloud is the red point, and the nine next most luminous
      ($M_{\rm lum}>8.6~\times~10^6\,$\msun) molecular clouds are blue. The
      kpc-scale estimate for \rco\ in NGC\,628 measured by EMPIRE is
      indicated with a dashed black line, and the average value for the
      cloud-scale measurements is represented by a dashed gray line. The
      \rco\ measurements in this plot use the peak brightness of the
      CO(1$-$0) and CO(2$-$1) lines within a $3\farcs2 \times 3\farcs2$ box
      centered on the cloud position.}
    \label{fig:r21-mass}
  \end{figure}}

\newcommand{\FigVelmapCOHa}{
  \begin{figure}
    \includegraphics[width=\linewidth]{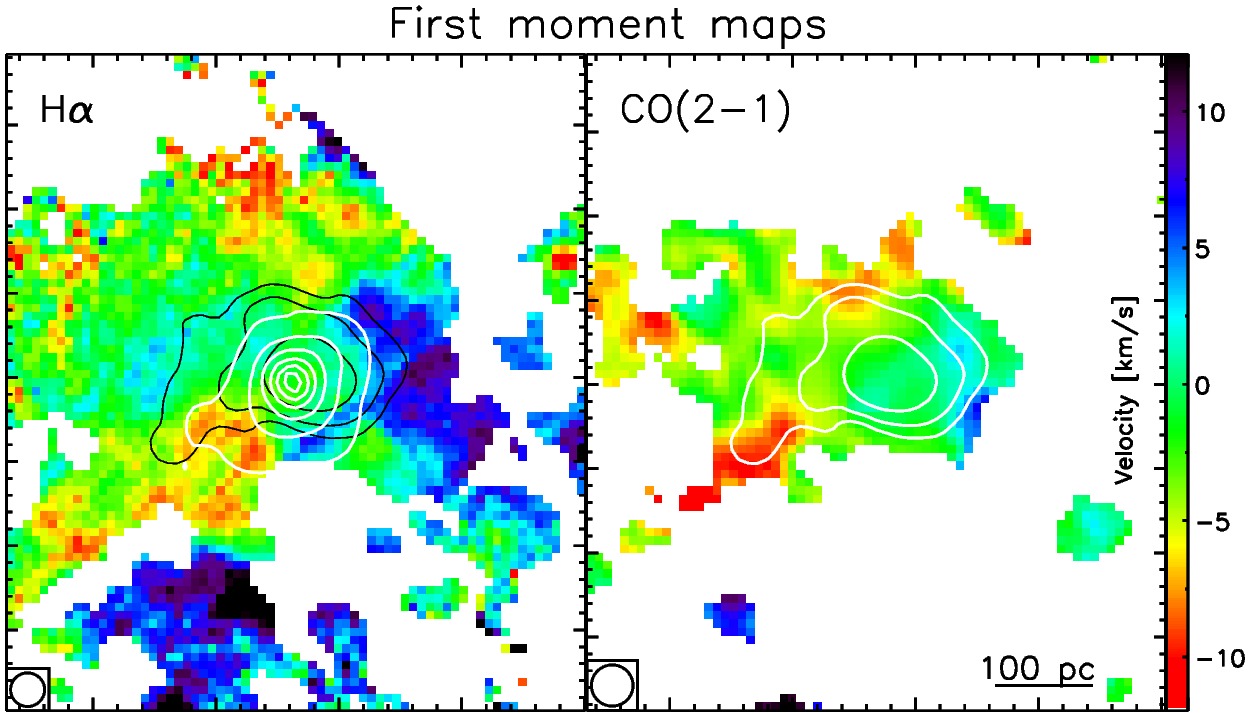}
    \caption{Maps of the velocity centroid of the H$\alpha$ and \co\
      emission for the region surrounding the headlight cloud. The maps are
      constructed after subtracting a model for the galaxy's rotation (Lang
      et al., in prep). The contours correspond to the H$\alpha$ and CO
      integrated emission: black (white) contours in the left (right) panel
      are the \co\ contours, white contours in the left panel are H$\alpha$
      contours. The H$\alpha$ seeing and the restored \co\ beam are
      indicated in the bottom-left part of each panel.}
    \label{fig:mom1COHa}
  \end{figure}}

\newcommand{\FigCDFHa}{
  \begin{figure}
    \centering %
    \includegraphics[width=\linewidth]{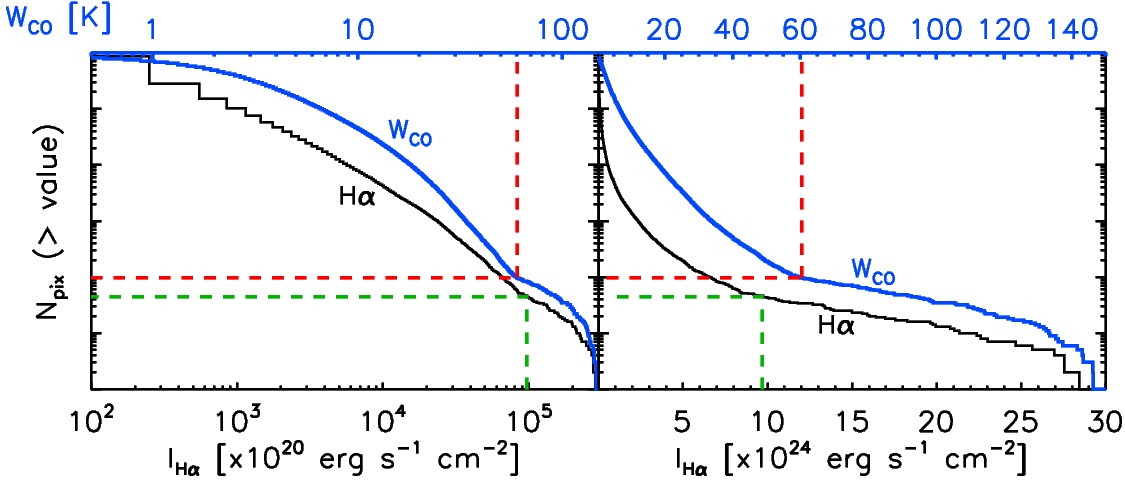}
    \caption{Number of pixels whose integrated H$\alpha$ intensity
      $I_{\rm H_{\alpha}}$ (black curve) and integrated \co\ intensity
      $W_{\rm CO}$ (blue curve) are above the value given in the bottom
      abscissa for $I_{\rm H\alpha}$ and in the top abscissa for
      $W_{\rm CO}$, for the entire galaxy. The left panel shows the full
      range of values in logarithmic scale while the right one zooms in
      linear scale. The red dashed lines are as in
      Fig.~\ref{fig:moments:full}. The green dashed lines show the values
      corresponding to 45 pixels (or a surface of about
      $1\farcs5\times1\farcs$5) in the H$\alpha$ map.}
    \label{fig:moments:cdfa}
  \end{figure}}

\newcommand{\FigRgalProps}{
  \begin{figure*}
    \centering
    \includegraphics[width=0.9\linewidth]{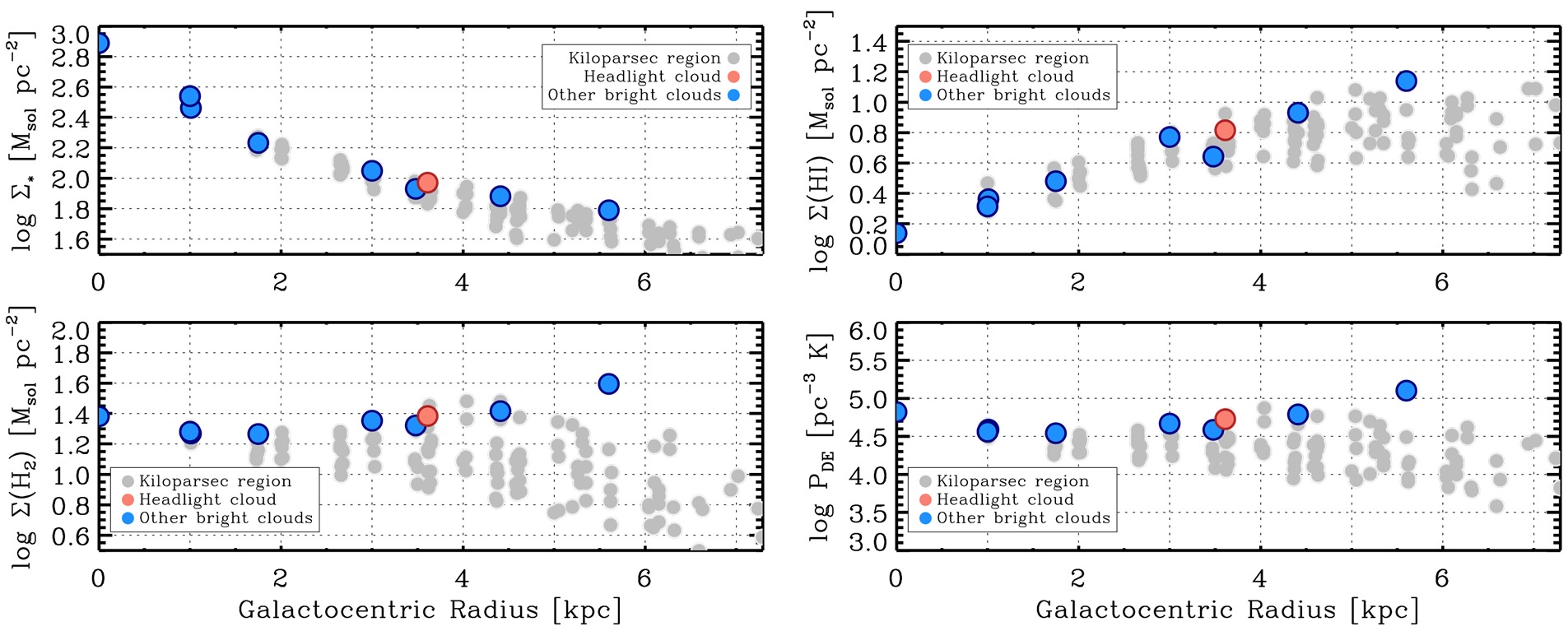}
    \caption{The headlight cloud in galactic context. Average surface
      density of stellar mass, atomic gas, molecular gas, and estimated
      dynamical equilibrium pressure in kiloparsec-sized apertures that
      cover the galaxy (from J. Sun et al. in preparation). Gray points
      show all apertures. Colored points show apertures that host one of
      the ten brightest clouds in the galaxy (red: the headlight cloud,
      blue: another bright cloud). The vertical spread of points at a fixed
      radius indicates azimuthal variations. The headlight cloud lies
      $\sim 3.5$~kpc from the galaxy center in an environment with
      unremarkable stellar mass or atomic gas content (see top
      row). However, this region does have among the highest molecular gas
      surface densities (bottom left) and dynamical equilibrium pressures
      (bottom right) anywhere in the galaxy. The headlight cloud and other
      bright cloud at similar radii are all associated with strong spiral
      features and so appear as the highest points at their radius. The
      fact that these clouds appear at large radius and comparatively low
      stellar surface density suggests that spiral structure and galactic
      dynamics (as opposed to only mean ISM pressure in a smooth disk) play
      a central role in setting the properties of these clouds.}
    \label{fig:rgal_props}
  \end{figure*}}

\newcommand{\FigCorotation}{
  \begin{figure*}
    \centering %
    \includegraphics[width=0.475\linewidth,angle=180]{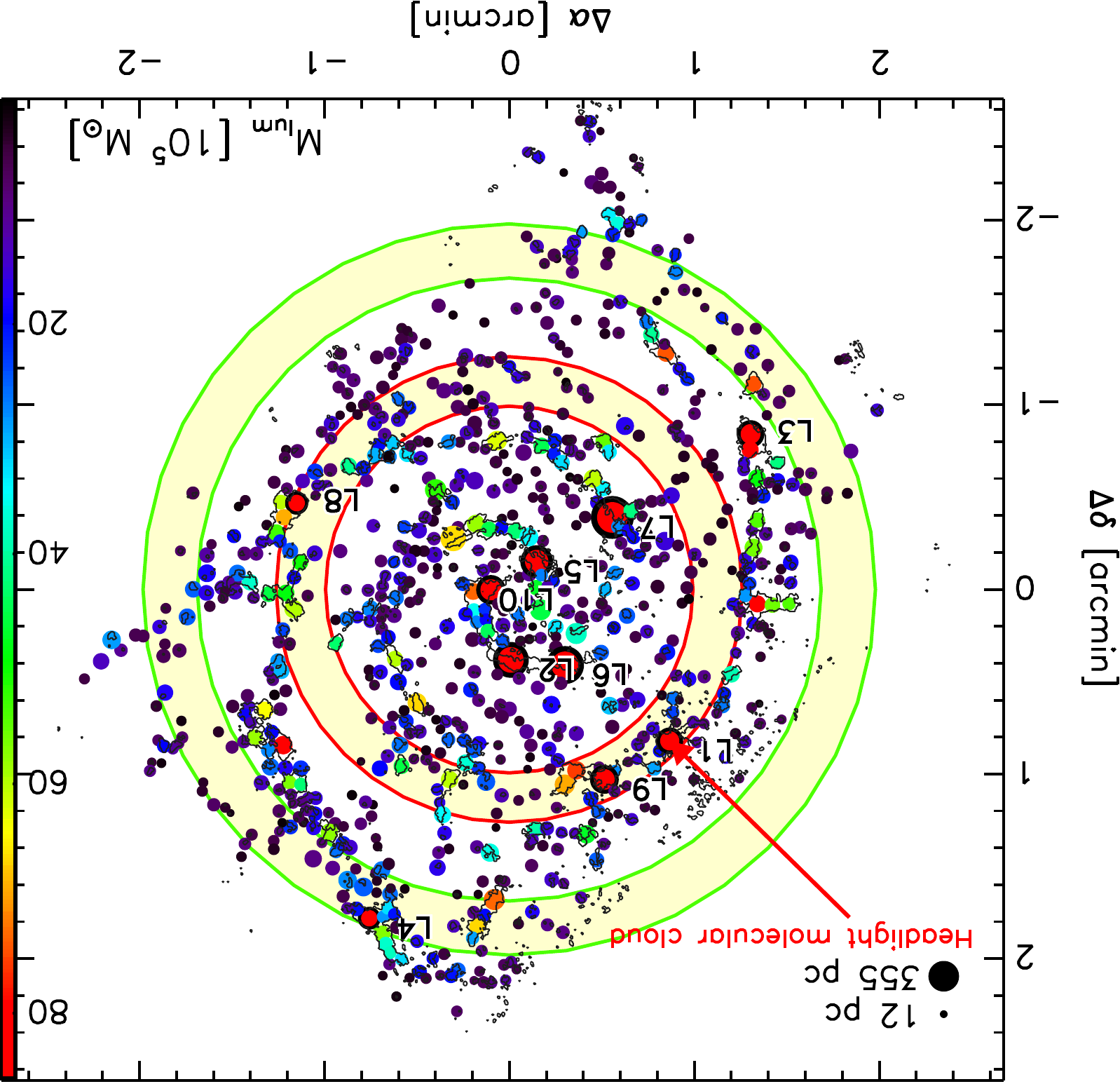}
    \includegraphics[width=0.475\linewidth,angle=180]{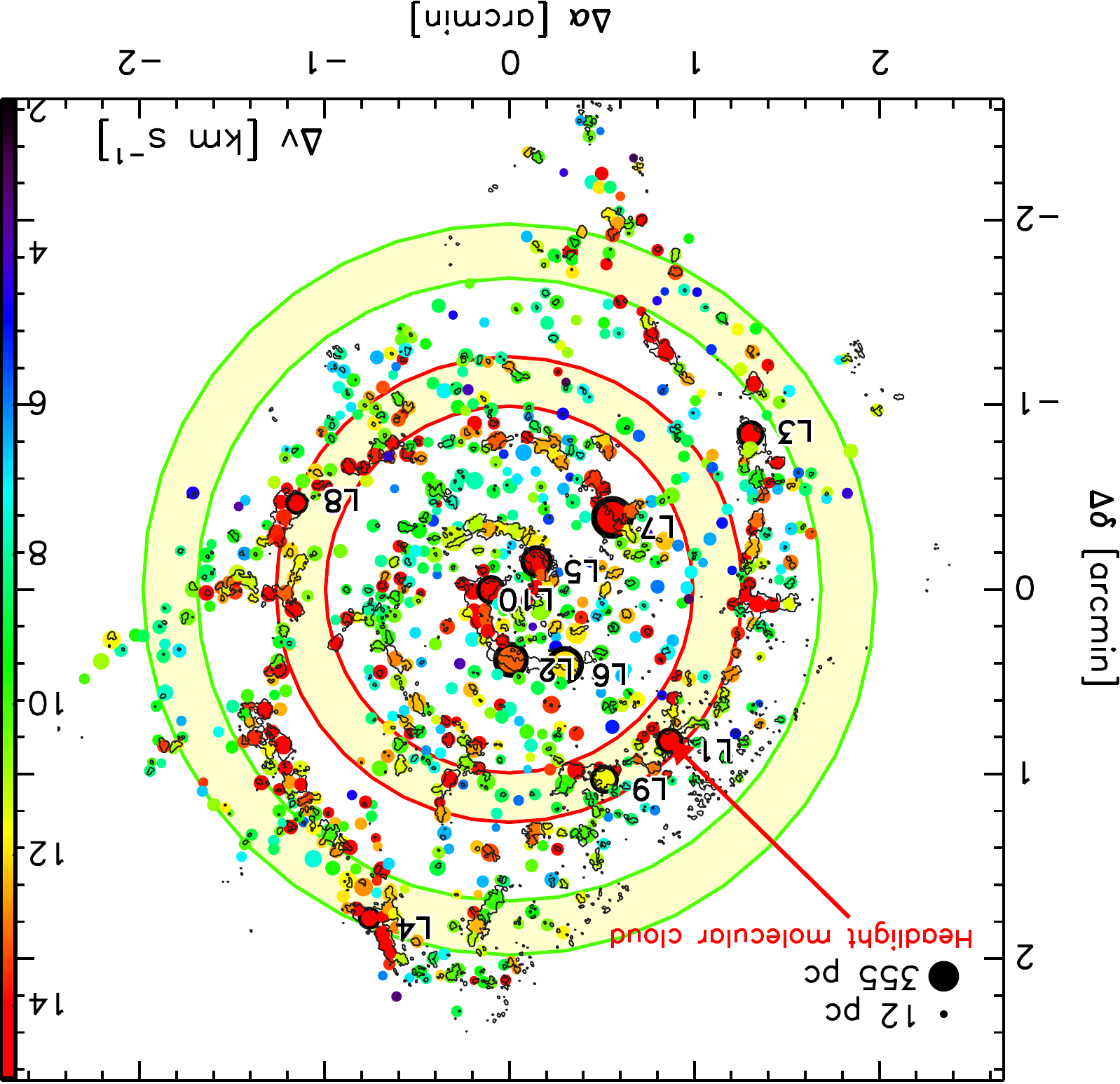}
    \includegraphics[width=0.475\linewidth,angle=180]{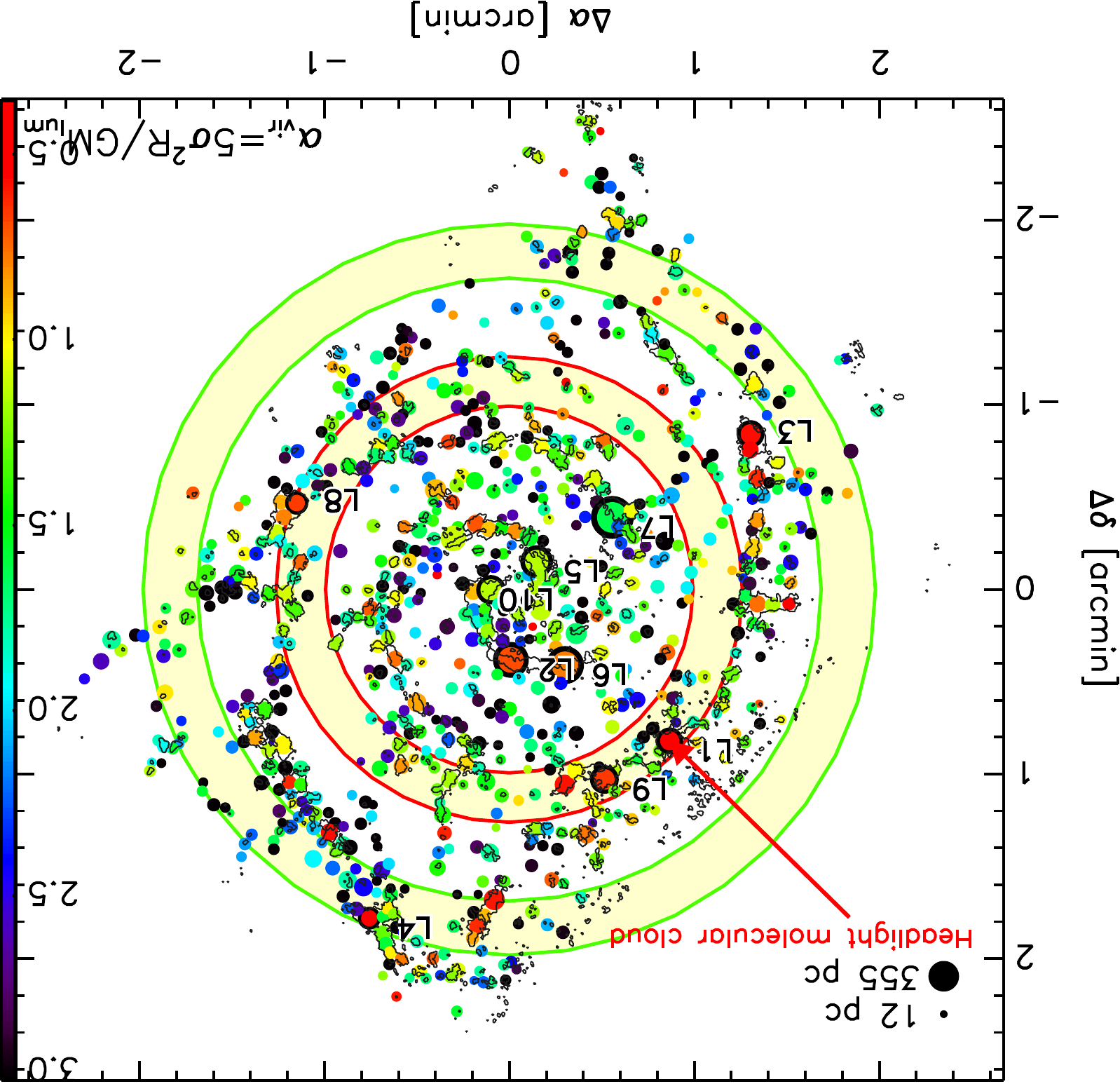}
    \includegraphics[width=0.475\linewidth,angle=180]{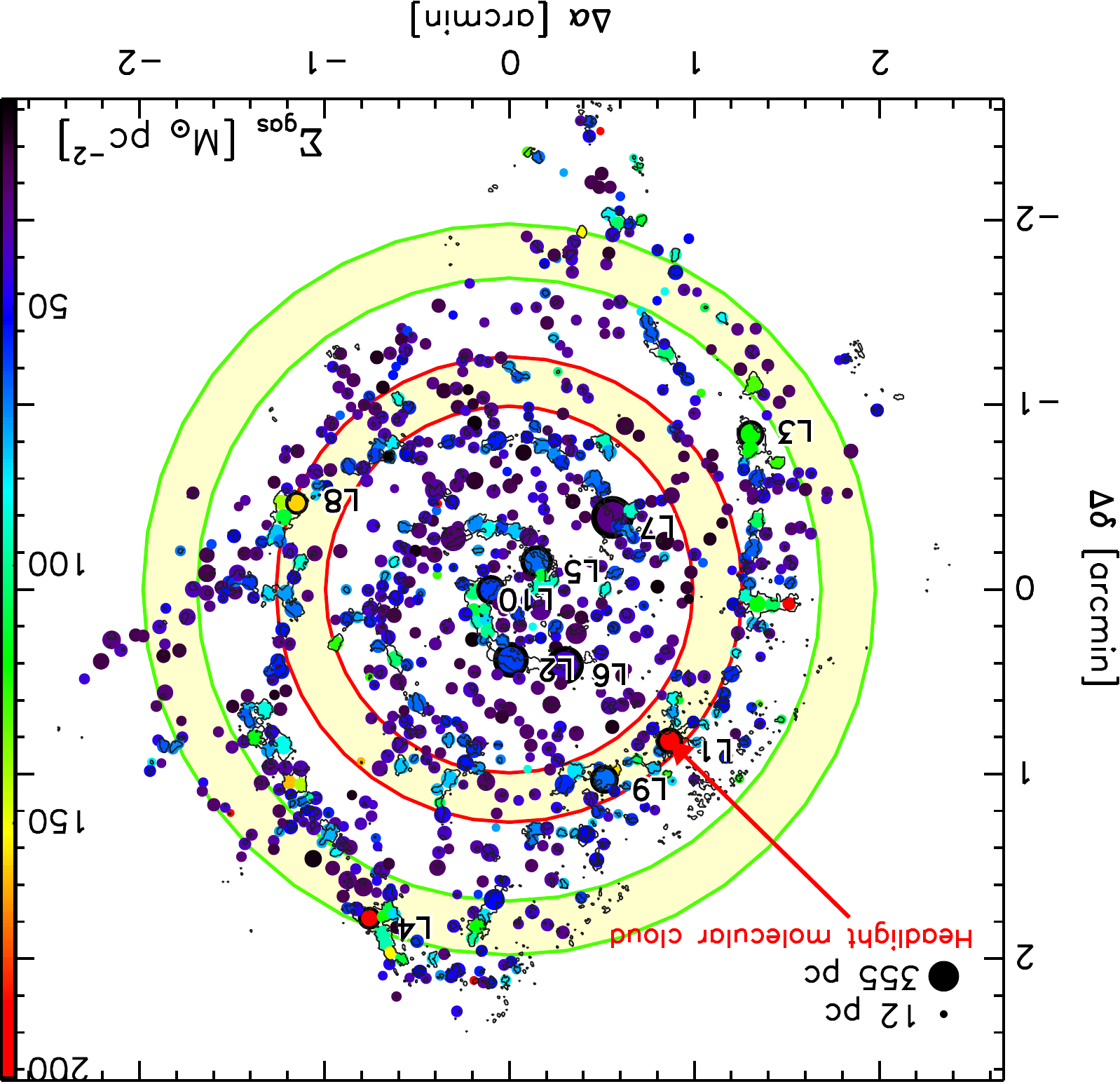}
    \caption{Position of molecular clouds in NGC\,628. From top-left to
      bottom-right, colors correspond to the CO luminous masses, CO
      velocity linewidth, alpha virial parameter, and gas surface density,
      truncated to 8.6$\times$10$^6$~\msun, 15~\kms, 3.0, and 200
      \msun\,pc$^{-2}$, respectively. Sizes of each symbol represent the
      physical size of the molecular clouds. In each panel we have marked
      the headlight molecular cloud (L1). The ten most massive molecular
      clouds are highlighted bold. The light yellow areas mark the position
      of corotation regions at 3.2\,kpc (red circles) and 5.1\,kpc (green
      circles). Contours correspond to the 5$\sigma$ level of the
        \co\ moment zero map. Offset positions are from the galactic
      center.}
    \label{fig:m74corotation}
  \end{figure*}}

\newcommand{\FigDynStructure}{
  \begin{figure}
    \centering %
    \includegraphics[width=0.95\linewidth]{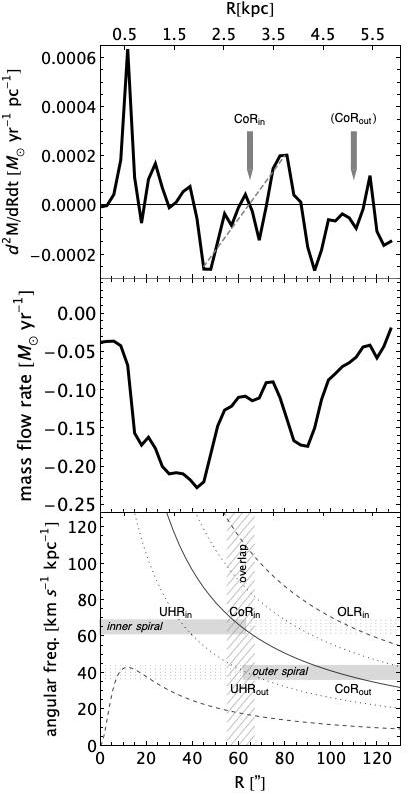}
    \caption{{\bf Top:} Profile of differential gas flow $d^2M/dR\,dt$
      driven by gravitational torques across NGC\,628.  The inner
      corotation radius (CoR$_{\rm in}$) marked by an arrow is located
      where torques and gas flows cross from negative (inward) to positive
      (outward) as illustrated by the short dashed segment. The second
      arrow marks the position of the outer corotation radius,
      CoR$_{\rm out}$, previously identified by \cite{cepa90}, which falls
      near the edge of the CO field-of-view. {\bf Middle:} Profile of
      integrated mass flow rates from the CoR$_{\rm out}$ inward to radius
      $R$.  Net flows are radially inward (negative) at all radii. The
      uncertainty on the inflow (not shown) is dominated by the large, 15''
      uncertainty in the position of the CoR$_{\rm out}$ near the map
      edge. {\bf Bottom:} Angular frequency curves in NGC\,628, $\Omega$
      (solid), $\Omega\pm\kappa/2$ (Dashed) and $\Omega\pm\kappa/4$
      (Dotted).  Positions of dynamical resonances and corotation radii are
      marked. The hashed vertical bar highlights the overlap
        between the CoR$_{\rm in}$ and the outer ultra-harmonic
        resonance.}
    \label{fig:dynstructure}
  \end{figure}}

\newcommand{\FigMultiLines}{
  \begin{figure*}
    \centering{} %
    \includegraphics[width=0.8\linewidth,angle=180]{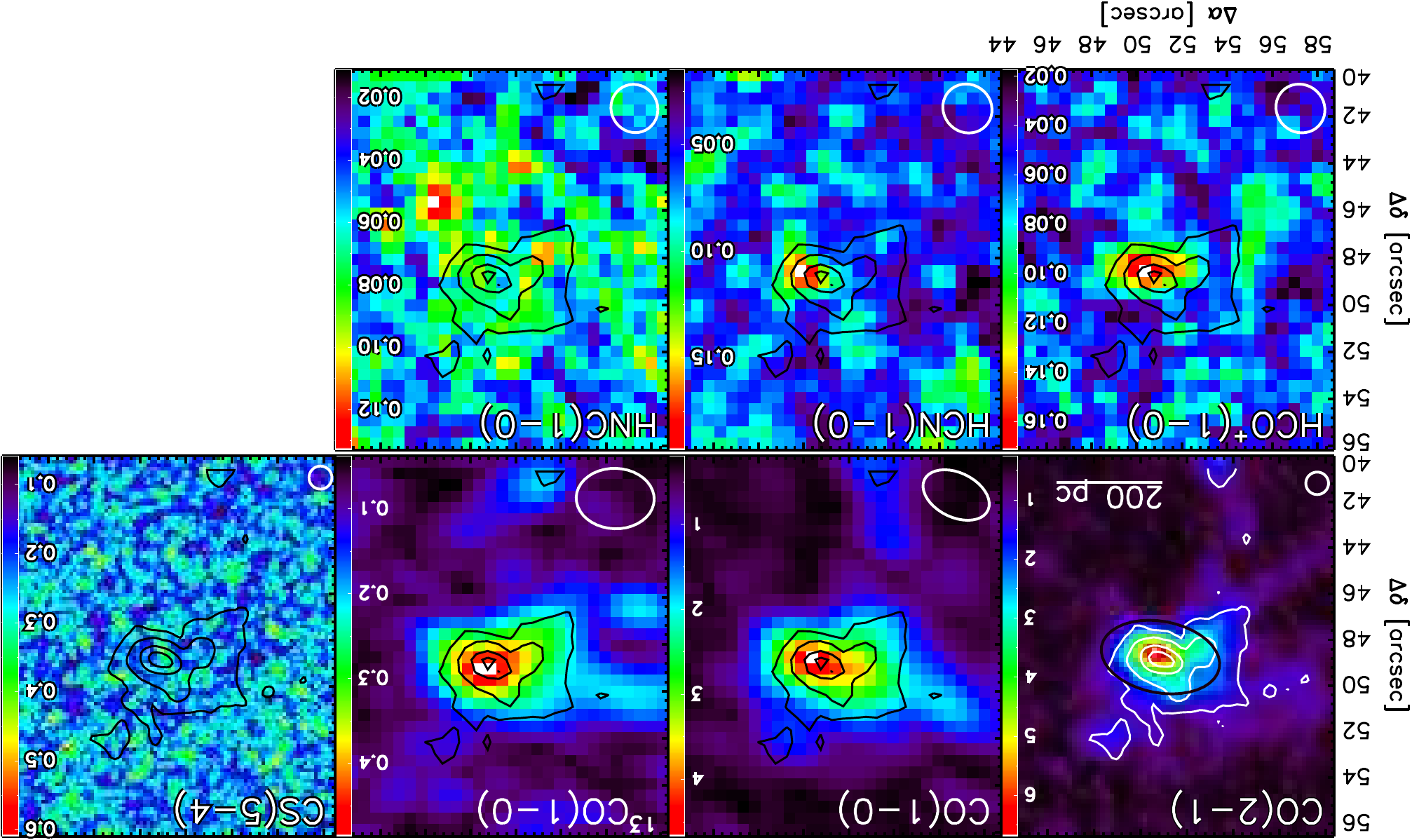}
    \caption{Peak temperature for the different gas tracers observed with
      ALMA. The bottom left inserts of each panel show the size of the
      synthesized beam.  Colorbars are in units of K. White/black contours
      in the first/other panels, respectively, correspond to the 20\%,
      40\%, 70\% and 90\% of the \co{} peak temperature. The top-left panel
      shows in a black ellipse the headlight cloud as identified by
      CPROPS. Offsets positions are from the galactic center.}
    \label{fig:cloudtracerstpeak}
  \end{figure*}
  \begin{figure*}
    \centering{} %
    \includegraphics[width=0.8\linewidth,angle=180]{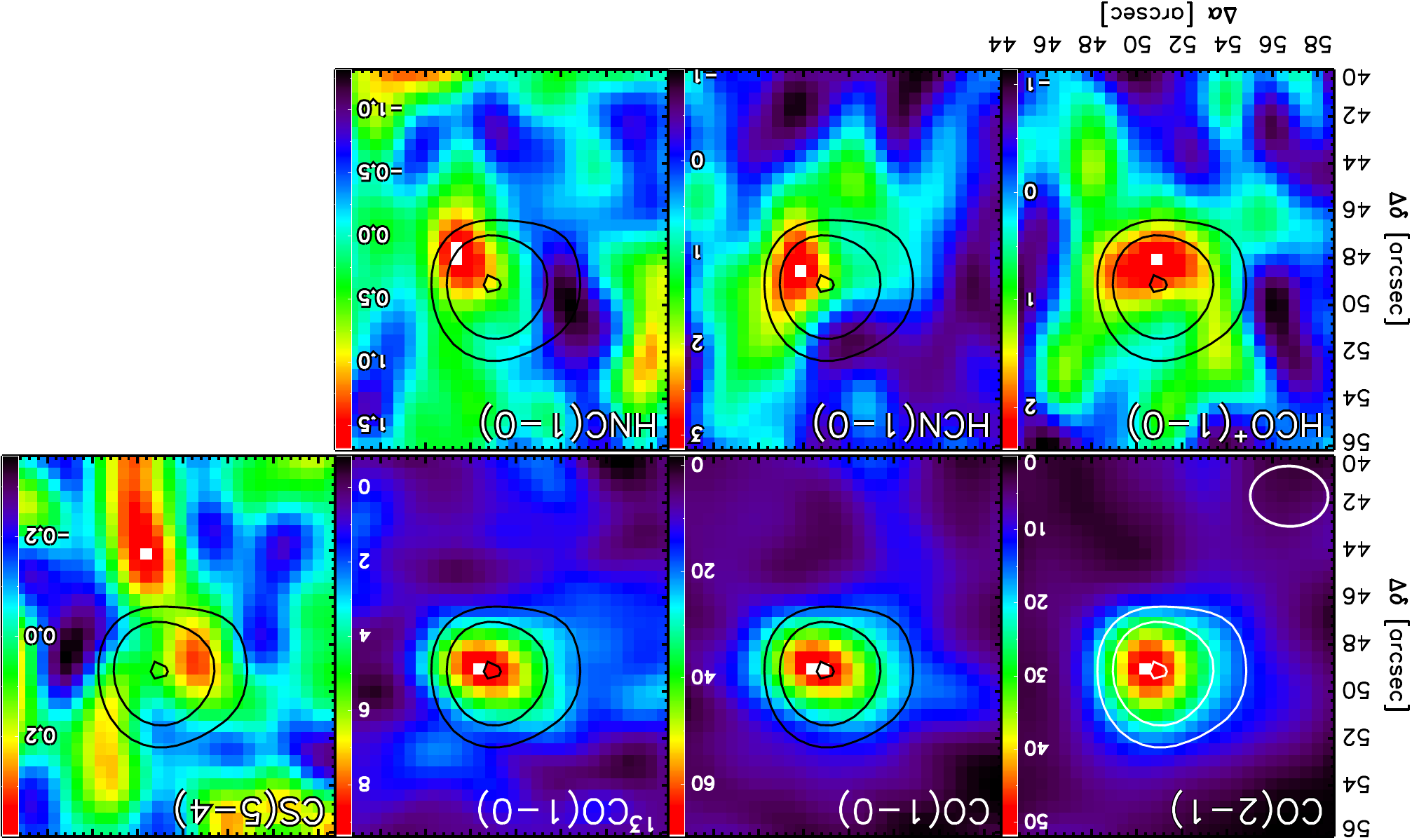}
    \caption{Integrated emission line for the different gas tracers
      observed with ALMA, convolved to the spatial resolution of
      $^{13}$CO(1$-$0).The bottom left inserts in the first panel shows the
      size of the synthesized beam. Emission for all tracers is integrated
      between 650 and 685~\kms, colorbars are in units of
      K~\kms. White/black contours in the first/other panels, respectively,
      correspond to 15$\sigma$, 25$\sigma$, 50$\sigma$ and 100$\sigma$,
      with $\sigma$= 1.01\,K\,\kms{} of the \co{} line. Offsets positions
      are from the galactic center.}
    \label{fig:cloudtracersIconv}
  \end{figure*}}
  
\newcommand{\TabData}{
  \begin{table}
    \begin{center}
      \caption{Observational parameters of the ALMA observations.}
      \label{tab:propobs}
      \begin{tabular}{cccccc}
        \hline
        \hline
        \noalign{\smallskip}
        \multirow{2}{*}{Line} & $\nu$ & noise & Restored   & $\Delta$v$_{\rm res}$ & 7m+\\
                              & GHz   &     K & beam  &         \kms & TP\\
        \hline
        \noalign{\smallskip}
        CS(5$-$4)        & 244.94 & 0.12$^{\rm a}$ & 1$\farcs$0 $\times$ 1$\farcs$0 & 5.0 &  no \\
        $^{12}$CO(2$-$1) & 230.54 & 0.17$^{\rm a}$ & 1$\farcs$0 $\times$ 1$\farcs$0 & 2.5 & yes\\
        $^{12}$CO(1$-$0) & 115.27 & 0.12  &  1$\farcs$9 $\times$ 3$\farcs$1  &  6.0 & yes \\
        $^{13}$CO(1$-$0) & 110.20 & 0.03  &  2$\farcs$6 $\times$ 3$\farcs$5  &  6.0 & yes \\
        HNC(1$-$0)       &  90.66 & 1.2   &  2$\farcs$0 $\times$ 2$\farcs$1  & 15.0 & no \\
        HCO$^+$(1$-$0)   &  89.19 & 1.1   &  2$\farcs$0 $\times$ 2$\farcs$2  & 15.0 & no \\
        HCN(1$-$0)       &  88.63 & 1.2   &  2$\farcs$1 $\times$ 2$\farcs$2  & 15.0 & no \\
        \hline
      \end{tabular}
    \end{center}
    $^{\rm a}$Sensitivity measured in the region that was not fully covered
    by the interferometric data, which is where the headlight cloud is
    located.  The sensitivity is 1.3 times better in the fully covered
    region.
  \end{table}}

\newcommand{\TabMassiveClouds}{
  \begin{table*}
    \centering{} %
    \caption{Physical properties of the ten most CO luminous clouds
      measured by CPROPS, sorted by decreasing luminous mass. Last line
      lists the parameters averaged in all molecular clouds in
      NGC\,628. From the first to the eleventh column: cloud
      identification, cloud coordinates in R.A. and Dec.,
      luminous CO mass, virial mass, radius, velocity linewidth, virial
      parameter, molecular gas mass surface density, \rco\ ratio, and
      measured distance to the closest corotation region. Column 12 lists
      the H$\alpha$ luminosity of the associated \ion{H}{ii} region
      estimated from MUSE observations from \citet{kreckel17}.}
    \label{tab:clouds}
    \resizebox{\linewidth}{!}{%
      \begin{tabular}{lccccccccccc}
        \hline
        \hline
        \noalign{\smallskip}
        & \multicolumn{10}{c}{Molecular gas} & \ion{H}{ii} region$^{\dagger}$ \\
        \hline
        \noalign{\smallskip}
        Cloud  & R.A.           & Dec.                            & M$_{\rm lum}$         & M$_{\rm vir}$         & Rad. & $\Delta_{v}$  & $\alpha_{\rm vir}$ & $\Sigma_{\rm mol}$       & R$_{21}$ & D$_{\rm corot}$ & L$_{\rm H\alpha}$ \\
        ID     & {\small h:m:s} & {\small $\degr:\arcmin:\farcs$} & {\small 10$^5$ \msun} & {\small 10$^5$ \msun} & pc   & {\small \kms} & --                 & {\small\msun\ pc$^{-2}$} & --       & kpc             & erg s$^{-1}$ \\
        \hline
        \noalign{\smallskip}
        L1$^{\star}$ & 01:36:45.328 & 15:47:48.48 &      204 &       91 &   184 &   16.3 &   0.5 &    192 & 0.70 &  0.00 &     6.3$\times$10$^{39}$ \\
        L2           & 01:36:41.757 & 15:47:22.17 &      149 &       92 &   280 &   13.2 &   0.7 &     61 & 0.61 &  1.71 &     5.5$\times$10$^{36}$ \\
        L3           & 01:36:47.144 & 15:46:08.77 &      142 &       73 &   193 &   14.2 &   0.6 &    121 & 0.71 &  0.37 &     7.1$\times$10$^{38}$ \\
        L4           & 01:36:38.574 & 15:48:46.11 &      139 &       68 &   141 &   16.0 &   0.5 &    223 & --$^{\dagger\dagger}$   &  0.00 &     8.2$\times$10$^{38}$ \\
        L5           & 01:36:42.337 & 15:46:50.22 &      127 &      131 &   246 &   16.9 &   1.2 &     66 & 0.60 &  2.18 &     2.6$\times$10$^{37}$ \\
        L6           & 01:36:42.974 & 15:47:23.85 &      118 &       83 &   302 &   12.1 &   0.8 &     41 & 0.60 &  1.35 &     8.6$\times$10$^{36}$ \\
        L7           & 01:36:44.038 & 15:46:35.88 &      118 &      172 &   355 &   16.1 &   1.6 &     30 & 0.56 &  0.87 &     4.2$\times$10$^{36}$ \\
        L8           & 01:36:36.942 & 15:46:31.21 &      105 &       63 &   145 &   15.2 &   0.7 &    159 & 0.61 &  0.00 &     6.6$\times$10$^{37}$ \\
        L9           & 01:36:43.859 & 15:48:01.02 &       88 &       52 &   204 &   11.6 &   0.7 &     67 & 0.57 &  0.00 &     1.3$\times$10$^{37}$ \\
        L10          & 01:36:41.309 & 15:46:59.40 &       86 &       89 &   212 &   15.0 &   1.2 &     61 & 0.63 &  2.50 &     7.7$\times$10$^{36}$ \\
        \hline
        NGC~628 Mean  & -- & -- &       14 &       22 &    91 &   10.3 &   2.1 &     56 &  0.54 & -- & -- \\
        \hline
      \end{tabular}}
    \tiny{} %
    \raggedright{} %
    $^{\star}$Cloud 1 corresponds to the NGC\,628 headlight.\\
    $^{\dagger}$Often there are multiple objects within a 150\,pc
    region. We have listed the closest \ion{H}{ii} region.\\
    $^{\dagger\dagger}$ Cloud L4 falls just outside the coverage of the CO(1$-$0) observations.
  \end{table*}}

\newcommand{\TabGaussianFits}{
  \begin{table}
    \centering %
    \caption{Best-fit parameters of the molecular line profiles. The line
      profiles are extracted at the position of the peak of the headlight
      cloud after convolution to a common angular resolution. The quoted
      uncertainties in the parameters correspond to the errors in the
      Gaussian fits, and the FWHM is reported as $\Delta v$. At 9.6\,Mpc,
      the $3\farc5 \times 2\farc6$ Gaussian beam of the matched resolution
      input data corresponds to a physical area of $\sim 22,300$\,pc$^2$.}
    \begin{tabular}{cccc}
      \hline
      \hline
      \noalign{\smallskip}
      & \multicolumn{3}{c}{At peak, convolved to $^{13}$CO beam} \\
      \hline
      Line & T$_{\rm Peak}$ & V$_{\rm LSR}$   & $\Delta$v \\
      & {\small[K]}    &  {\small[\kms]} &  {\small[\kms]} \\
      \hline
      \noalign{\smallskip}
      CO(2$-$1)        & 2.52$\pm$0.02 &  668.3$\pm$0.1 & 19.5$\pm$0.2\\
      CO(1$-$0)        & 3.64$\pm$0.11 &  668.3$\pm$0.3 & 18.1$\pm$0.6\\
      $^{13}$CO(1$-$0) & 0.51$\pm$0.02 &  668.2$\pm$0.4 & 17.7$\pm$0.9\\ 
      HCO$^{+}$(1$-$0) & 0.10$\pm$0.01 &  666.8$\pm$1.8 & 20.4$\pm$3.0\\
      HCN(1$-$0)       & 0.09$\pm$0.02 &  660.4$\pm$2.7 & 32.2$\pm$6.5\\
      \hline
    \end{tabular}
    \label{tab:fitprof}
  \end{table}}

\newcommand{\TabLineRatHeadlight}{
  \begin{table}
    \begin{center}
      \caption{Line ratios at the peak of the headlight cloud. The line
        ratios are calculated using (i) the peak brightness and (ii) the
        integrated intensity of each emission line, as determined by a
        Gaussian fit to the line profile (see Table~\ref{tab:fitprof}). The
        line profiles are extracted at the position of the peak of the
        headlight cloud after convolution to a common angular resolution.}
      \label{tab:linerat}
      \begin{tabular}{cccccc}
        \hline
        \hline
        \noalign{\smallskip}
        Ratio & Peak & Integrated \\
        \hline
        \noalign{\smallskip}
        \medskip
        $\dfrac{\rm CO(2-1)}{\rm CO(1-0)}$           & 0.70  & 0.75  \\
        \medskip
        $\dfrac{\rm ^{12}CO(1-0)}{\rm ^{13}CO(1-0)}$ & 7.2   & 7.3   \\
        \medskip
        $\dfrac{\rm HCO^{+}(1-0)}{\rm HCN(1-0)}$     & 1.1   & 0.7   \\
        \medskip
        $\dfrac{\rm HCO^+(1-0)}{\rm ^{12}CO(1-0)}$   & 0.028 & 0.031 \\
        \medskip
        $ \dfrac{\rm HCN(1-0)}{\rm ^{12}CO(1-0)}$    & 0.025 & 0.043 \\
        \medskip
        $ \dfrac{\rm HCN(1-0)}{\rm ^{13}CO(1-0)}$    & 0.2   & 0.3   \\
        \hline
      \end{tabular}
    \end{center}
  \end{table}}
  
\newcommand{\TabLineRatGals}{
  \begin{table*}
    \begin{center}
      \caption{$^{13}$CO/$^{12}$CO(1$-$0), \rco, HCO$^+$/HCN and
        HCN/CO(1$-$0) line ratios for the different sources displayed in
        Figure~\ref{fig:rat-gals}. For nearby galaxies from the EMPIRE
        survey, the average ratios are given for entire galaxies, disks and
        centers. The physical sizes used to derive these ratios are also
        listed.}
      \label{tab:rat-gals}
      \resizebox{\linewidth}{!}{%
        \begin{tabular}{cccccccccccccccccc}
          \hline
          \hline
          \noalign{\smallskip}
          \multirow{2}{*}{Source} & \multicolumn{4}{c}{$^{12}$CO/$^{13}$CO(1$-$0)} & \multicolumn{4}{c}{CO(2$-$1)/CO(1$-$0)} & \multicolumn{4}{c}{HCO$^+$/HCN(1$-$0)} & \multicolumn{4}{c}{HCN/CO(1$-$0)}  & \multirow{2}{*}{Ref.$^\dagger$}\\
          \cline{2-17}
                                  & Size & \multicolumn{3}{c}{Mean ratio} & Size & \multicolumn{3}{c}{Mean ratio} & Size & \multicolumn{3}{c}{Mean ratio}& Size & \multicolumn{3}{c}{Mean ratio}& \\
          \hline
          \noalign{\smallskip}
          Headlight &   140 pc & \multicolumn{3}{c}{7.20} & 140 pc & \multicolumn{3}{c}{0.70} & 140 pc & \multicolumn{3}{c}{1.0} & 140 pc & \multicolumn{3}{c}{0.03}  & This work \\   
          Orion~B   &   6.5 pc & \multicolumn{3}{c}{6.50} & 6.5 pc & \multicolumn{3}{c}{0.75} & 6.5 pc & \multicolumn{3}{c}{1.10} & 6.5 pc & \multicolumn{3}{c}{0.92}  & S94,P17 \\ 
          30~Dor-10 &     8 pc & \multicolumn{3}{c}{6.47} &  10 pc & \multicolumn{3}{c}{1.07} & 1.5 pc$^{\dagger\dagger}$ & \multicolumn{3}{c}{5} & 1.5 pc$^{\dagger\dagger}$ & \multicolumn{3}{c}{0.14}  & P12,J98,A14\\ 
          \hline
          \multicolumn{2}{l}{EMPIRE galaxies} & Gal. & Disk & Cen.&   & Gal. & Disk & Cen.&  & Gal. & Disk & Cen.&    & Gal. & Disk & Cen.& C18, JD\\ 
          \cline{3-5}\cline{7-9}\cline{11-13}\cline{15-17}
          \noalign{\smallskip}
          NGC~6946   & 0.9~kpc&  13.8 &  13.5 &  15.2 & 1.1~kpc&  0.65 &  0.64 &  0.68 & 1.1~kpc&  1.06 &  1.19 &  0.87 & 1.1~kpc&  0.73 &  0.50 &  1.14 & \\ 
          NGC~5194   & 1.0~kpc&  10.3 &   9.9 &   8.2 & 1.2~kpc&  0.73 &  0.81 &  0.80 & 1.2~kpc&  0.75 &  0.97 &  0.65 & 1.2~kpc&  1.11 &  1.17 &  2.26 & \\ 
          NGC~5055   & 1.0~kpc&   9.3 &   8.5 &   7.2 & 1.3~kpc&  0.59 &  0.59 &  0.68 & 1.3~kpc&  0.33 &  0.35 &  0.90 & 1.3~kpc&  1.19 &  1.07 &  0.71 & \\ 
          NGC~2903   & 1.2~kpc&  10.3 &  11.0 &  12.3 & 1.4~kpc&  0.63 &  0.64 &  0.68 & 1.4~kpc&  0.53 &  0.42 &  0.72 & 1.4~kpc&  0.92 &  0.83 &  1.38 & \\ 
          NGC~3627   & 1.2~kpc&  12.0 &  12.2 &  15.2 & 1.5~kpc&  0.48 &  0.47 &  0.48 & 1.5~kpc&  0.90 &  0.87 &  0.91 & 1.5~kpc&  0.76 &  0.59 &  0.70 & \\ 
          NGC~0628   & 1.3~kpc&  13.2 &  13.5 &   9.9 & 1.5~kpc&  0.61 &  0.64 &  0.57 & 1.5~kpc&  0.55 &  0.47 &  0.89 & 1.5~kpc&  0.54 &  0.72 &  0.63 & \\ 
          NGC~3184   & 1.5~kpc&  10.2 &  10.7 &  11.2 & 1.9~kpc&  0.53 &  0.53 &  0.52 & 1.9~kpc&  0.52 &  0.55 &  0.77 & 1.9~kpc&  0.42 &  0.39 &  0.62 & \\ 
          NGC~4321   & 1.9~kpc&  10.3 &  10.3 &  10.9 & 2.3~kpc&  0.57 &  0.56 &  0.62 & 2.3~kpc&  0.90 &  1.03 &  0.75 & 2.3~kpc&  1.14 &  0.84 &  1.57 & \\ 
          NGC~4254   & 1.9~kpc&  11.0 &  10.5 &   8.4 & 2.3~kpc&  0.74 &  0.74 &  0.75 & 2.3~kpc&  1.19 &  1.23 &  0.79 & 2.3~kpc&  0.75 &  0.70 &  1.23 & \\ 
          \hline
        \end{tabular}}
    \end{center}
    $^{\dagger}$S94: \citet{sakamoto94}; P17: \citet{pety17}; P12:
    \citet{pineda12}; J98: \citet{johansson98}; A14: \citet{anderson14};
    C18: \citet{cormier18}; JD: \citet{jimenez-donaire19}.\\
    $^{\dagger\dagger}$ The lines were observed at an angular resolution
    corresponding to 29\,pc.
  \end{table*}}

\newcommand{\TabHalphaProp}{
  \begin{table}
    \centering{} %
    \caption{Estimated physical properties of the \ion{H}{ii} region and
      young stellar population embedded in the headlight cloud in
      NGC\,628.}
    \label{tab:halpha-prop}
    \resizebox{\linewidth}{!}{%
      \begin{tabular}{ccc}
        \hline
        \hline
        \noalign{\smallskip}
        Name & Symbol & Value \\
        \hline
        \noalign{\smallskip}
        H$\alpha$ luminosity             & L$_{\rm H\alpha}$ & $6.3\times 10^{39}$~erg~s$^{-1}$ \\
        H$\alpha$ region radius          & ---               & 142\,pc \\
        H$\alpha$ line width             & $\Delta$v         & $\sim$50~\kms \\
        H$\alpha$ equivalent width       & W$_{\rm H\alpha}$ &  517~\AA \\
        Ionizing photon production rate  & N$_{\rm ion}$     & $4.6\times 10^{51}$ phot. s$^{-1}$ \\
        Young star mass                  & M$_{\rm cl}$      & $3\times 10^5$~\msun \\
        Young star bolometric luminosity & L$_{\rm cl}$      & $4\times 10^8$~L$_{\odot}$ \\
        Young star typical age           & ---               & $2-4$~Myr \\
        \hline
      \end{tabular}}
  \end{table}}

\begin{document}

\title{The headlight cloud in NGC\,628: \\
  An extreme giant molecular cloud in a typical galaxy disk}

\author{Cinthya~N.~Herrera\inst{1}\fnmsep\thanks{herrera@iram.fr} %
  \and Jérôme Pety\inst{1,2} %
  \and Annie Hughes\inst{3,4} %
  \and Sharon E. Meidt\inst{5,6} %
  \and Kathryn Kreckel\inst{5} %
  \and Miguel Querejeta\inst{7,8} %
  \and Toshiki Saito\inst{5} %
  \and Philipp Lang\inst{5} %
  \and María Jesús Jiménez-Donaire\inst{9} %
  \and Ismael Pessa\inst{5} %
  \and Diane Cormier\inst{10} %
  \and Antonio Usero\inst{9} %
  \and Kazimierz Sliwa\inst{5} %
  \and Christopher Faesi\inst{5} %
  \and Guillermo A. Blanc\inst{11,12} %
  \and Frank Bigiel\inst{13} %
  \and Mélanie Chevance\inst{14} %
  \and Daniel A. Dale\inst{15} %
  \and Kathryn Grasha\inst{16} %
  \and Simon C. O. Glover\inst{17} %
  \and Alexander P. S. Hygate\inst{5,14} %
  \and J.~M.~Diederik Kruijssen\inst{14} %
  \and Adam K. Leroy\inst{18} %
  \and Erik Rosolowsky\inst{19} %
  \and Eva Schinnerer\inst{5} %
  \and Andreas Schruba\inst{20} %
  \and Jiayi Sun\inst{18} %
  \and Dyas Utomo\inst{18} %
}

\institute{Institut de Radioastronomie Millim\'etrique, 300 rue de la
  Piscine, 38406 Saint-Martin-d'H\`eres Cedex 
  \and Sorbonne Universit\'e, Observatoire de Paris, Universit\'e PSL,
  CNRS, LERMA, 75014, Paris, France 
  \and CNRS, IRAP, 9 av. du Colonel Roche, BP 44346, 31028 Toulouse cedex
  4, France 
  \and Universit\'e de Toulouse, UPS-OMP, IRAP, 31028 Toulouse cedex 4,
  France 
  \and Max Planck Institut f\"ur Astronomie, K\"onigstuhl 17, D-69117
  Heidelberg, Germany 
  \and Sterrenkundig Observatorium, Universiteit Gent, Krijgslaan 281 S9,
  B-9000 Gent, Belgium 
  \and European Southern Observatory, Karl-Schwarzschild-Str. 2, D-85748
  Garching, Germany 
  \and Observatorio Astron{\'o}mico Nacional (IGN), C/ Alfonso XII 3, 28014
  Madrid, Spain 
  \and Harvard \& Smithsonian | Center for Astrophysics, 60 Garden
  St. MS-78, 02138, Cambridge, USA 
  \and The Observatories of the Carnegie Institution for Science, 813 Santa
  Barbara St., Pasadena, CA, 91101, USA 
  \and Departamento de Astronomía, Universidad de Chile, Camino del
  Observatorio 1515, Las Condes, Santiago, Chile 
  \and AIM, CEA, CNRS, Universit\'e Paris-Saclay, Universit\'e Paris
  Diderot, Sorbonne Paris Cit\'e, F-91191 Gif-sur-Yvette, France 
  \and Argelander-Institut f\"ur Astronomie, Universit\"at Bonn, Auf dem
  H\"ugel 71, 53121 Bonn, Germany 
  \and Astronomisches Rechen-Institut, Zentrum f\"{u}r Astronomie der
  Universit\"{a}t Heidelberg, M\"{o}nchhofstra\ss e 12-14, 69120
  Heidelberg, Germany 
  \and Department of Physics \& Astronomy, University of Wyoming, Laramie,
  WY 82071 USA 
  \and Research School of Astronomy and Astrophysics, Australian National
  University, Canberra, ACT 2601 Australia 
  \and Institut f\"ur Theoretische Astrophysik, Zentrum f\"ur Astronomie
  der Universit\"at Heidelberg, Albert-Ueberle-Strasse 2, 69120
  Heidelberg 
  \and Department of Astronomy, The Ohio State University, 140 West 18th
  Ave, Columbus, OH 43210, USA 
  \and Department of Physics, University of Alberta, Edmonton, AB T6G 2E1,
  Canada 
  \and Max-Planck-Institut f\"ur extraterrestrische Physik,
  Giessenbachstra\ss e 1, D-85748 Garching, Germany 
}

\authorrunning{Herrera et al.} %

\titlerunning{The headlight cloud in NGC\,628}

\date{\today}
 
\abstract
{Cloud-scale surveys of molecular gas reveal the link between giant
  molecular cloud properties and star formation across a range of galactic
  environments. Cloud populations in galaxy disks are considered to be
  representative of the `normal' star formation process, while galaxy
  centers tend to harbour denser gas that exhibits more extreme star
  formation.  At high resolution, however, molecular clouds with
  exceptional gas properties and star formation activity may also be
  observed in normal disk environments. In this paper, we study the
  brightest cloud traced in \co\ emission in the disk of nearby spiral
  galaxy NGC 628.}
{We characterize the properties of the molecular and ionized gas that is
  spatially coincident with an extremely bright \ion{H}{ii} region in the
  context of the NGC\,628 galactic environment.  We investigate how
  feedback and large-scale processes influence the properties of the
  molecular gas in this region.}
{High resolution ALMA observations of \co\ and CO(1$-$0) emission are used
  to characterize the mass and dynamical state of the ``headlight''
  molecular cloud.  The characteristics of this cloud are compared to the
  typical properties of molecular clouds in NGC\,628. A simple large
  velocity gradient (LVG) analysis incorporating additional ALMA
  observations of $^{13}$CO(1$-$0), HCO$^{+}$(1$-$0) and HCN(1$-$0)
  emission is used to constrain the beam-diluted density and temperature of
  the molecular gas. We analyze the MUSE spectrum using Starburst99 to
  characterize the young stellar population associated with the \ion{H}{ii}
  region.}
{The unusually bright headlight cloud is massive
  ($1 - 2 \times 10^7$~\msun), with a beam-diluted density of
  $n_{\rm H_2}=5\times10^4$~cm$^{-3}$ based on LVG modeling. It has a low
  virial parameter, suggesting that the CO emission associated with this
  cloud may be overluminous due to heating by the \ion{H}{ii} region. A
  young ($2-4$\,Myr) stellar population with mass $3\times10^{5}$~\msun{}
  is associated.}
{We argue that the headlight cloud is currently being destroyed by feedback
  from young massive stars. Due to the cloud's large mass, this phase of
  the cloud's evolution is long enough for the impact of feedback on the
  excitation of the gas to be observed. The high mass of the headlight
  cloud may be related to its location at a spiral co-rotation radius,
  where gas experiences reduced galactic shear compared to other regions of
  the disk, and receives a sustained inflow of gas that can promote the
  cloud's mass growth.}
  

\maketitle{} %
 
\section{Introduction}
\label{ss:intro}

NGC\,628\ (M74) is a nearby \citep[$d=9.6$\,Mpc,][]{kreckel17}, almost
face-on \citep[$i\sim9\degr$,][]{blanc13}, type SAc grand-design spiral
galaxy with a stellar mass of $M_{\star} = 1.5{-}2 \times 10^{10}\,\msun$
\citep{querejeta15,leroy19} and moderate global star
formation rate (SFR) of $\sim2\,\msun\,$yr$^{-1}$ \citep[][]{sanchez11}. A
well-known Messier object, NGC\,628 is one of about ten galaxies that has
been observed by nearly all recent major surveys of the interstellar gas
and dust in nearby galaxies, including THINGS, HERACLES, SINGS, KINGFISH
and EMPIRE
\citep[][]{walter08,leroy09,kennicutt03,kennicutt11,bigiel16}. NGC\,628 is
likewise a popular target for observational studies of ionized gas and star
formation in the local Universe, with existing wide-field WFPC3/ACS
observations by the Hubble Space Telescope (HST) Legacy Extragalactic UV Survey (LEGUS) program \citep[][]{calzetti15} that enable a
detailed characterisation of the population of stellar clusters and
associations in NGC\,628, as well as high-resolution, wide-field optical
Integral Field Unit (IFU) imaging by the VENGA survey and CFHT/SITELLE
\citep[][]{blanc13,rousseau-nepton18}.

Using the Multi Unit Spectroscopic Explorer (MUSE) optical IFU instrument
on the Very Large Telescope (VLT), \citet{kreckel16,kreckel18} recently
noted an extremely bright \ion{H}{ii} region in the outer part of the
galaxy \citep[it is also visible, but uncommented on, in earlier H$\alpha$
mapping, e.g.,][]{ferguson98,lelievre00}. The source stands out as a
bright, compact peak in the MUSE H$\alpha$ map. It is two orders of
magnitude brighter than the mean \ion{H}{ii} region in the galaxy
population and twice as bright as the next most luminous source identified
by \citet[][see their Fig.~1]{kreckel18}. The \ion{H}{ii} region has an
equivalent radius of 142\,pc, a velocity dispersion of 50\,\kms, and an
H$\alpha$ luminosity of $6.3\times10^{39}\,$erg\,s$^{-1}$, which
corresponds to a local SFR of $\sim 0.034$\,\msun\,yr$^{-1}$, adopting the
calibration of \citet{kennicutt12}. This bright \ion{H}{ii} region is
associated with bright, compact infrared emission in \textit{Spitzer} and
\textit{Herschel} maps \citep[e.g., see][]{aniano12}. It appears as the
brightest spot in the galaxy in maps of WISE 12$\mu$m and 22$\mu$m emission
\citep{leroy19}. The region thus exhibits the classic
observational signatures of a large population of luminous young stars that
are still associated with a large reservoir of interstellar gas and dust.

The association with gas is borne out by our new   Atacama  Large  Millime-
ter/submillimeter Array (ALMA) CO observations. We
observed NGC\,628 at $1'' \sim 50$\,pc resolution in \co\ as part of a
PHANGS-ALMA survey\footnote{PHANGS, Physics at High Angular resolution in
  Nearby GalaxieS, is an international collaboration aiming to understand
  the interplay of the small-scale physics of gas and star formation with
  galactic structure and evolution. \texttt{http://www.phangs.org}.}  (PI:
E. Schinnerer; co-PIs: A.~Hughes, A.~K.~Leroy, A.~Schruba,
E.~Rosolowsky). These NGC\,628 CO maps have already appeared in
\citet{leroy16}, \citet{sun18}, \citet{kreckel18}, and \citet{utomo18}. The
ALMA data reveal an exceptionally bright CO peak spatially coincident with
the \ion{H}{ii} region \citep[the association is particularly striking in
Fig.~1 of][]{kreckel18}. While the source was visible in earlier CO maps
\citep[e.g.,][]{wakker95,helfer03,leroy09,rebolledo15}, ALMA shows it to be
stunningly compact and bright. This source is thus a compact,
dust-enshrouded collection of many massive, young stars still associated
with what appears to be the most massive molecular cloud in NGC\,628. This
object is even more remarkable because of its location at large
galactocentric radius, which makes it distinct from the gas-rich, intensely
star-forming regions that are commonly identified in galaxy centers.

A preliminary census of the disk environments in the PHANGS-ALMA galaxy
sample suggests that this type of object is relatively rare, but not
unique.  To understand the internal and external factors that can influence
the formation and evolution of such massive molecular clouds and their
extraordinary \ion{H}{ii} regions away from galaxy centers, a thorough
examination of local physical conditions in the star-forming gas is
essential.  In this paper, we present the first such study, focusing on the
CO-bright cloud in NGC\,628 as a prototype.  We name this object the
\textit{headlight cloud,} because it appears as a bright spot in an
otherwise almost dark (unsaturated) map of the H$\alpha$ and CO emission.

Several factors make the headlight cloud the ideal candidate for this
preliminary study. As noted above, NGC\,628 is one of a limited (but
growing) set of targets with information for a suite of millimetre and
optical emission lines that can be used to constrain the physical
properties of the molecular gas and the young stellar population.  Relevant
global galaxy properties (e.g. SFR) and the galaxy-scale gas dynamics in
NGC\,628 can be precisely characterized using the wealth of
multi-wavelength imaging data. NGC\,628 is also part of
LEGUS~\citep{calzetti15}. Studies using the LEGUS cluster catalog present a
detailed quantitative overview of NGC\,628's cluster population, and the
ability of NGC 628's gas disk to form clusters and regulate their
evolution~\citep{adamo17,grasha17,ryon17,grasha15}. Nevertheless, the
region hosting the headlight cloud lies just outside the coverage of the
final multi-wavelength LEGUS mosaic.

Our goal in this paper is to measure the physical conditions in the
molecular gas of the headlight cloud and to quantitatively describe the
associated star formation activity.  By comparing the measurements for the
headlight cloud to the rest of the galaxy, we aim to build a picture of
what may have prompted the growth of such a massive cloud and its extreme
star formation event. The paper is structured as
follows. Section~\ref{ss:obs} presents the observational data that we
use. Section~\ref{ss:gas_properties} describes the properties of the
molecular gas. In Section~\ref{ss:feedback}, we discuss the stellar
formation and feedback in the headlight cloud based on the MUSE data. In
Section~\ref{ss:galaxy_environment} we investigate the galactic environment
of the headlight cloud. Section~\ref{ss:discussion} discusses the
results. Our conclusions are summarized in Section~\ref{ss:conclusions}.

\section{Observations}
\label{ss:obs}

NGC\,628 was observed with ALMA in Chile during ALMA's Cycle~1 (ID: 2012.1.00650.S, PI:
E. Schinnerer) and Cycle~2 (ID: 2013.1.00532.S, PI: E. Schinnerer). While
the MUSE and CO~(2$-$1) data have previously appeared elsewhere (see
Sect.~\ref{ss:intro}), the 3~mm line observations are presented here for
the first time.

\subsection{1\,mm lines}

The observational strategy, calibration and imaging of the interferometric
data, array combination, and data product delivery of the PHANGS ALMA \co\
data are described in a forthcoming dedicated PHANGS-ALMA survey paper
(Leroy et al., in prep.). In Section~\ref{ss:b6obs} and~\ref{ss:b6reduction}, we
briefly summarize the key steps for completeness.  The present paper is the
appropriate citation for the reduction of the PHANGS-ALMA total power data,
which we present in detail in appendix~\ref{app:tp} and summarize
below. The strategy that we have used to reduce the \co\ total power data
for NGC\,628 has been adopted as the basis for the PHANGS-ALMA total power
processing pipeline.

\subsubsection{Observations}
\label{ss:b6obs}

ALMA Band 6 observations were obtained during Cycle 1 to image the emission
of $^{12}$\co\ (\co\ hereafter), CS(5$-$4) and the 1\,mm continuum. The
covered frequency ranges in the source frame were 229.6-230.5~GHz and
231.0-232.8~GHz in the lower sideband (LSB), and 244.0-245.9~GHz and
246.7-246.8~GHz in the upper sideband (USB). The targeted field of view was
a rectangular area of 240\farc$\times$175$\farc$ centered on the galaxy
nucleus.

In order to recover all the spatial scales of the \co\ emission, 12-m, 7-m,
and Total Power observations were performed. On-The-Fly observations with
three single-dish 12-m antennas delivered the Total Power. The off-source
position was chosen at the offset $[240\farc,420\farc]$ from the phase
center located at $\alpha=01^h 36 ^m 41\fs72$,
$\delta=+15\degr 46\arcmin 59\farcs3$ in the equatorial J2000
frame. Mosaics with a total of 149 and 95 pointings were observed with 27
to 38 antennas of the main 12-m array and with 7 to 9 antennas of the 7-m
array, respectively. Eight, ten and twenty-four execution blocks were
observed for the 12-m, 7-m, and Total Power arrays, respectively.

\subsubsection{Reduction and Imaging}
\label{ss:b6reduction}

\TabData{} %

The \co\ observations were part of the pilot program for PHANGS-ALMA. Leroy
et al. (in prep.) describe the selection, reduction, and imaging of the
data in detail.

In summary, interferometric calibration followed the recipes provided in
the ALMA reduction scripts. Visual inspection of the different calibration
steps (bandpass, phase, amplitude, and flux) showed that the scripts
yielded a satisfactory calibration. For these data, calibration was
performed within the Common Astronomy Software Application (CASA), versions
4.2.2 and 4.2.1 for the 12-m and 7-m array observations,
respectively. We used the task \texttt{statwt} on calibrated
  science data to ensure that the weights were estimated in a consistent
  way for 7-m and 12-m data reduced with different versions of
  CASA.\footnote{See the CASA/ALMA documentation at
    \url{https://casaguides.nrao.edu/index.php/DataWeightsAndCombination}.}

We imaged and post-processed the data using CASA version 5.4.0. The
procedure is broadly as follows. We spectrally regrid all data onto a
common velocity frame with channel width $\sim 2.5$~km~s$^{-1}$. Then we
combine and jointly image all 12-m and 7-m data together. We deconvolved
the data in two stages. First, we carry out a multi-scale clean using the
CASA task \texttt{tclean}. We used a broad clean mask and cleaned until the
residual maps has a maximum signal-to-noise ratio of $4$. Then, we created
a more restrictive clean mask and restarted the cleaning using a single
scale clean at the highest angular resolution in order to ensure
  that all point sources were completely deconvolved. We ran this single
scale clean until the flux in the model met a convergence criteria related
to the fractional change in the model flux during successive clean
deconvolution cycles.

The Total Power data were reduced with CASA version 4.5.3 following a
procedure described in Appendix~\ref{app:tp}. Briefly, atmospheric and flux
calibration as well as the data gridding into a position-position-velocity
cube followed the recipes provided by ALMA. However, we did not
  use the baseline recipe delivered by ALMA. Instead, we fitted each
individual spectrum before gridding with a polynomial of order 1 on the
same baseline window for all the spectra. This fixed baseline
window covered the $[0,590~\kms]$ and $[750,800~\kms]$ velocity range,
while the ALMA pipeline was trying to automatically adjust the
velocity windows as a function of position. While this idea is appealing
in principle, it is difficult to implement without \textit{a priori}
information on the source kinematics. In practice, it was biasing the
baselining process by confusing low brightness signal with baseline
noise. In our scheme, we used our \textit{a priori} information on the
location of the galaxy signal in the velocity space.

After imaging, we converted the units of the cube to Kelvin, primary beam
corrected the data. The synthesized beam of the 12m+7m
interferometric image is convolved with a Gaussian kernel to reach an
angular full width at half maximum (FWHM) resolution of 1\farcs0 (47\,pc in linear scale at a distance
of 9.6\,Mpc). We then projected the single dish data onto the
astrometric grid of the interferometric image and used the CASA task
\texttt{feather} to combine the two data sets.

We list the properties of the final cubes in Table~\ref{tab:propobs}. Note
that the sensitivity of our final data cube is not homogeneous because a
fraction of the 12-m and 7-m execution blocks did not cover the full
field-of-view. A band of $\sim$40\arcsec width towards the North-Western
edge thus has a noise 1.3 times higher than the remainder of the
rectangular field-of-view. The headlight source lies within this band.

For CS(5$-$4), a simple single-resolution clean was applied to the 12-m
data within the same mask as for the \co\ emission. The 7-m and Total Power
data were not added in this case because no emission was detected in these
datasets. The deconvolved image was also convolved with a Gaussian
to reach an angular resolution of 1\farcs0.

\subsection{3\,mm lines}

\subsubsection{Observations}

\FigMomentsFull{} %

Additional observations were acquired in Cycle 2 to image the emission from
the $J=1-0$ transition of $^{12}$CO (CO(1$-$0) hereafter), $^{13}$CO,
C$^{18}$O, HCO$^+$, HCN, and HNC, as well as the CS(2$-$1) and HNCO
5(0,5)$-$4(0,4) lines. Three band 3 frequency setups at 90, 110 and 115 GHz
were required to cover all these emission lines. The first setup at 90 GHz
covers the $87.9-91.4$ GHz (LSB) and $99.6-103.2$ GHz (USB) frequency
range. The second setup covers the $96-99.4$ GHz (LSB) and $108-111.7$ GHz
(USB) range, and the third setup covers the $100.1-103.8$ GHz (LSB) and
$112.3-115.5$ GHz (USB) range. The 110~GHz and 115~GHz setups target the CO
lines that show both extended and compact emission. Thus, 7-m array and
Total Power observations were performed in addition to the 12-m array
observations in order to correctly recover the flux at all scales.
Short-spacing observations for the 90~GHz set-up are available for
NGC\,628, but we chose not to incorporate them into the data cubes that we
analyze here, because the 90~GHz setup targets lines mostly show compact
emission structures.

The field-of-view covered by our observations at 90, 110, and 115~GHz was
232\farc$\times$170$\farc$, 240\farc$\times$150$\farc$, and
200\farc$\times$150$\farc$, respectively. Between 31 and 38 12-m antennas
and between 9 and 10 7-m antennas were used during the different
observations. The Total Power observations used simultaneously either 2 or
3 antennas. The same OFF as for the Cycle 1 was used during the On-The-Fly
observations.

\subsubsection{Reduction and Imaging}

\FigZoom{} %

The data reduction recipes used for the 3\,mm lines are much
simpler than the ones used to reduce the \co. This is due to the fact
that the 3\,mm lines were reduced much before the recipes for the large
program converged. In view of the satisfactory results obtained at
relative low signal-to-noise ratio, we did not try to refine them. This
section describes these simple recipes.

Data calibration for the Cycle 2 band 3 data was performed in CASA,
versions 4.2.1 and 4.3.1 for the interferometric data, and 4.4 and 4.5 for
the Total Power data. Interferometric calibration followed the recipes
provided in the ALMA reduction scripts, as visual inspection showed that
the scripts yielded a satisfactory calibration.

The CO and $^{13}$CO (1$-$0) data were then spectrally smoothed to
6\,km\,s$^{-1}$. A signal mask was created from a CO(1$-$0) total power
cube smoothed to an angular resolution of $60''$. The mask was created
using the CPROPS signal detection algorithm, i.e. emission regions with a
signal-to-noise ratio greater than 3 were identified and then extended to
include contiguous pixels with a signal-to-noise ratio greater than 2. This
mask was used to guide the deconvolution of the CO and $^{13}$CO (1$-$0)
data.

A single-scale clean algorithm, as coded in the CASA task \texttt{tclean},
was applied to the 12\,m+7\,m data up to the point where the residual
maximum fell below 1\,mJy. The total power cube was used as an initial
model. The resulting cleaned cube was then feathered with the total power
cube.

The fainter lines were imaged at coarser spatial and spectral resolution
(see Table~\ref{tab:propobs} for details). Total power cubes were not used
during the deconvolution (neither as models, nor for short-spacing
correction) in these cases as the signal was barely or not at all detected
in the single-dish observations.

\subsection{Complementary MUSE data}
\label{sec:muse}

We use MUSE observations to characterize the ionized gas and stellar
content associated with the headlight cloud. MUSE is an optical IFU on
the VLT in Paranal, Chile. \citet{kreckel16,kreckel18} present a detailed explanation of the
strategy and data reduction of the MUSE observations of NGC\,628. We
summarize a few key aspects here.

The observations cover a wavelength range from 4800 to 9300\,\AA, which
includes the H$\alpha$ and H$\beta$ emission lines and a large region of
the optical stellar continuum. The velocity resolution is about
150~\kms. In the mode that we used, the field-of-view of the IFU is
$1\arcmin \times 1\arcmin$ with $0\farcs2$ pixels. We paneled multiple
fields-of-view to cover the whole inner part of the galaxy
\citep[see][]{kreckel18}. The seeing of the observations was $\sim1$\arcsec
($\sim47$\,pc in linear scale), with an astrometric accuracy of $0\farcs2$.

The H$\alpha$ emission line was obtained using the LZIFU pipeline
\citep{ho16}, which simultaneously fits and subtracts the stellar
continuum. All line fluxes quoted in this paper have been corrected for
extinction. The emission line reddening has been inferred by comparing the
observed H$\alpha$ to H$\beta$ line ratio to an intrinsic value of 2.86
(for case B recombination and an electron temperature of 10,000\,K) under
the assumption of a \citet{Fitzpatrick99} extinction curve with $R_V=3.1$.
Typical values for E(B-V) are $\sim 0.5$ including for the headlight cloud.

\section{Molecular gas in the headlight cloud}
\label{ss:gas_properties}

\subsection{The \co{} emission}
\label{ss:co:headlight}

\TabMassiveClouds{} %

The top row of Fig.~\ref{fig:moments:full} presents ALMA maps of the \co{}
peak brightness and integrated intensity across NGC\,628. The headlight
cloud is located within an outer spiral arm, at an offset of $[47'',51'']$
and a radial distance of 3.2\,kpc from the galaxy center.  The cloud is
visible as the brightest peak in both panels of
Fig.~\ref{fig:moments:full}.

In the bottom row of Fig.~\ref{fig:moments:full}, we plot pixel-wise
cumulative distribution functions (CDFs) of the \co{} peak brightness
$T_{\rm CO}$ and the integrated intensity $W_{\rm CO}$ across the ALMA
field-of-view. The CDF is a one-point statistic that quantifies how the
emission in each map is distributed between low- and high-brightness
regions \citep[for detailed discussion of CO distribution functions we
refer the reader to ][]{hughes13b}.

The headlight cloud stands out in both CDFs. For both $T_{\rm CO}$ and
$W_{\rm CO}$, the distributions clearly show a change in slope for the 100
brightest pixels in the map, which we indicate with a red line. The sense
is that above some high threshold, there is more bright CO emission than
one would predict based on the rest of the galaxy. That is, the slope of
the distribution function becomes flatter. All of these unusually bright
pixels above either threshold belong to a region surrounding the headlight
cloud.

This bright emission is notable for both its shape and location. We
constructed equivalent CDFs for the peak brightness and integrated
intensity maps of the CO(1$-$0) emission measured by the PAWS survey of M51
\citep{schinnerer13,pety13} and the MAGMA survey of the Large Magellanic Cloud (LMC) \citep{wong11},
and of \co\ emission measured by the IRAM\,30-m survey of M33
\citep{druard14}, finding no similar features \citep[see also ][ but note
that this paper uses the probability distribution function rather than CDF formalism]{hughes13b}.  In the
mass distribution functions plotted by \citet[][including NGC\,628]{sun18},
bumps and features at high brightness are almost always associated with
galaxy centers or dynamical environments like stellar bars. The headlight
cloud lies in the outer part of a spiral arm, not associated with either
the galaxy center or a stellar bar.

In Fig.~\ref{fig:moments:zoom}, we zoom in on the headlight cloud. These
maps show the \co{} peak brightness, integrated intensity,
centroid velocity and FWHM in a large region
around the cloud (left) and in the immediate vicinity of the cloud
(right). They show highly concentrated CO emission, with a compact, bright
peak ($T_{\rm peak}> 6\,$K over 1 square arcsecond $\sim
2,000$\,pc$^2$). This bright peak is surrounded by a more extended
component with $T_{\rm peak}> 1\,$K over $\sim 50$ square arcseconds or
$\sim 100,000$\,pc$^{2}$.

Morphologically, the cloud appears to be linked to other material in the
spiral arm by four filaments. A fifth filament extends southwards from the
cloud into the interarm region. The velocity field within the cloud, as
traced by the centroid velocity, is relatively smooth, while the gas in the
southern filament appears shifted to higher velocities compared to the
cloud material.

The gas velocity dispersion in the cloud is high, and increases towards the
cloud center: within $\sim2$ arcseconds of the intensity peak, the FWHM
linewidths are $>15\,\kms$ with a maximum value of $24\,\kms$.  Typical
\co{} FWHM linewidths in the surrounding spiral arm are $\le10\,\kms$ (see
Fig.~\ref{fig:moments:zoom} top right panels).

\subsection{GMC properties in NGC\,628}
\label{ss:cprops}

\FigLarson{} %

Catalogs of giant molecular clouds (GMCs) for all the PHANGS-ALMA galaxies,
including NGC\,628, have been constructed and will be presented in a
forthcoming paper (Rosolowsky et al., in prep). These catalogs are
generated using the CPROPS algorithm \citep{rosolowsky06} to segment the
\co\ data cubes into individual molecular clouds and then measure the
radius, linewidth, luminosity and other structural properties of each
cloud.

\subsubsection{The headlight cloud}
\label{ss:cprops:headlight}

\FigStacks{} %

The cataloged radius and (FWHM) linewidth of the headlight cloud are
184\,pc and 16.3\,\kms{} respectively. The measured radius of
184\,pc has been deconvolved by the beam size, extrapolated to
correct for sensitivity and blending effects, and then converted to the
radius convention defined by \citet{solomon87}, which implies multiplying
the rms size by $1.91$~\citep[for details see,][]{rosolowsky06}. For a
Gaussian shape, this implies a cloud FWHM size of $\sim 220$\,pc.

The headlight cloud is the most luminous out of 850 clouds in the
catalog, with a CO luminosity mass of
$M_{\rm lum}=\alpha_{\rm CO}\,L_{\rm CO} = 2.0\times10^{7}$\,\msun. This
mass assumes a CO(2$-$1)-to-H$_2$ conversion factor
$\alpha_{\rm CO} = 6.2$~\msun~(K\,\kms\,pc$^2$)$^{-1}$, which is the
Galactic CO(1$-$0) to H$_2$ conversion factor,
4.3~\msun~(K\,\kms\,pc$^2$)$^{-1}$ \citep{bolatto13}, divided by
0.69, i.e., the CO(2$-$1)/CO(1$-$0) ratio measured in the
headlight cloud (see Sect.~\ref{sec:overluminous}). This \rco\ value is
close to the nearby galaxy canonical value of 0.7
\citep[e.g.,][]{leroy09,leroy13, saintonge17} and to the mean ratio
measured on kpc-scales in NGC\,628 by the recent EMPIRE survey \citep[][see
also Table~\ref{tab:rat-gals}]{jimenez-donaire19}.

Treating the geometry as a three-dimensional Gaussian, the measured mass
and radius of the headlight cloud imply an H$_2$ density of
${\rm n_{H_2}\sim20}$~cm$^{-3}$. This is considerably lower than the
critical density of CO(2$-$1) emission, even allowing for line trapping,
and implies significant clumping or substructure within the cloud. The bulk
physical parameters of the headlight cloud correspond to free-fall time and
Mach number typical of massive giant molecular clouds. Using a mean density
${\rm n_{H_2}\sim20}$~cm$^{-3}$, the free-fall time
$t_{\rm ff}=\sqrt{3 \pi /(32G\rho)}$ and crossing time
$t_{\rm cross}=R/\Delta v$ are 8 and 11~Myr respectively. Assuming a gas
temperature of $\sim$20~K (see Section~\ref{ss:lvg}), the measured
linewidth corresponds to a Mach number of $\sim 42$.

\subsubsection{Comparison with other luminous clouds in NGC\,628}

\FigLinesLVG{} %

To place these values in context, Table~\ref{tab:clouds} reports the
physical properties (mass, radius, virial parameter) of the ten most
luminous GMCs in NGC\,628, as well as the mean value for NGC\,628's entire
cataloged GMC population. We compare the headlight cloud to the full GMC
population, highlighting these massive clouds, in Fig.~\ref{fig:larson}.

The 184\,pc radius of the headlight cloud is large but does not
clearly distinguish the cloud from other massive clouds in NGC\,628. The
global linewidth of the headlight cloud (16.3\,km\,s$^{-1}$) is
also large but similar to that of the other high-mass GMCs in NGC\,628. As
a result, in the bottom left panel of Fig.~\ref{fig:larson}, the headlight
cloud clusters with the other massive clouds at the high end of the line
width-size relation.

In contrast, the headlight cloud does stand out in surface density. The
bottom right panel of Fig.~\ref{fig:larson} shows that the mass surface
density of the headlight cloud, estimated from the luminous mass, is higher
than any other massive cloud but one and among the highest in the
galaxy. This value, $\Sigma_{\rm CO}\sim$
  190\,${\rm M}_{\odot}\,{\rm pc}^{-2}$, is roughly three times greater
than the average value for all GMCs in NGC\,628.

The most exceptional property of the CO(2$-$1) emission in the headlight
cloud is its peak brightness, i.e., the intensity of the brightest pixel in
the cloud at the $\sim1\arcsec$ resolution of our PHANGS-ALMA CO(2$-$1)
data. In the top row of Fig.~\ref{fig:larson}, the headlight cloud (in red)
clearly separates from the other massive clouds (in blue), with a peak
brightness of almost $7$~K. This is roughly twice the peak brightness of
the other massive clouds and the highest value found for any cloud in the
galaxy.

A consequence of the headlight cloud's high surface density and high
brightness is that the cloud appears tightly bound. The virial parameter of
the headlight cloud, estimated from the ratio between the cloud's CO
luminous mass and the virial mass $\alpha_{\rm vir}$=5$\sigma_{\rm v}^2$R/GM$_{\rm lum}$ is low, 0.5. This
is low compared to both most other clouds in the galaxy and the value
expected for virialized ($\alpha_{\rm vir}=1$) or marginally bound
($\alpha_{\rm vir}=2$) GMCs.  We note that the low virial parameter of the
headlight cloud (and many of the other massive clouds) appears to be part
of a systematic trend in NGC\,628 for the GMC virial parameters to decrease
with increasing peak brightness (top left panel of Fig.~\ref{fig:larson})
and $M_{\rm lum}$ (not shown). This trend needs to be interpreted with
care: the scatter at low brightness is partly due to the impact of marginal
resolution on these measurements, while CO brightness also enters in the
denominator of the virial parameter via the definition of $M_{\rm
  lum}$. Nevertheless, simulations do predict mass- and
environment-dependent variations in cloud virialization
\citep[e.g.][]{federrath12}, and we plan to investigate these trends in
more detail using the full PHANGS-ALMA sample in a future paper.

Figure~\ref{fig:larson} shows that the headlight cloud has a large size and
velocity dispersion like the other massive clouds in NGC\,628, but an
exceptionally bright core. This implies a unique profile for the cloud
within the NGC\,628 GMC population. Figure~\ref{fig:stacks} examines this
profile in more detail. The top left panel plots the CO(2$-$1) integrated
intensity as a function of distance away from the brightest pixel in the
cloud for each of the 10 most massive GMCs in NGC\,628. The top right panel
shows the same, but normalizing all clouds by the integrated intensity at
the cloud centre. The bottom row of Fig.~\ref{fig:stacks} is similar, but
here we plot radial profiles of the CO(2$-$1) line width instead of
integrated intensity. We use the second moment to parameterize the cloud
line width in Fig.~\ref{fig:stacks}, but a similar result is obtained using
other line width diagnostics.

Figure~\ref{fig:stacks} shows that the headlight cloud has the brightest
core among all the massive clouds. The core is surrounded by an extended
envelope of lower intensity emission -- leading to the large size estimated
by CPROPS -- but the top right panel shows that in relative terms, the
headlight cloud has the most compact profile of all massive clouds in
NGC\,628. The velocity profile of the headlight cloud is also distinctive,
with an enhanced central line width and a steeper decrease in the line
width between the cloud's center and edge.

Figures~\ref{fig:larson} and \ref{fig:stacks} thus tend to reinforce the
visual impression from Fig.~\ref{fig:moments:full} that the CO emission
associated with the headlight cloud has exceptional properties.  After
integrating over the extent of the CPROPS-identified cloud, the headlight
cloud resembles other massive clouds in NGC\,628 in terms of size, line
width, and mass. Yet looking more closely, it stands out due to its compact
profile and enhanced central line width. Below, we show that the headlight
cloud's compact CO-bright core is coincident with a bright \ion{H}{ii}
region. This, combined with its apparently strong self gravity, indicates a
massive star-forming region caught in the early stages of forming stars,
and perhaps a potential birth site of massive clusters.

\subsection{Other molecular emission lines}
\label{ss:other:tracers}

In addition to \co{}, we observed NGC\,628 in CO(1$-$0), $^{13}$CO(1$-$0),
CS(5$-$4), HCO$^{+}$(1$-$0), HCN(1$-$0) and HNC(1$-$0). The ratios among
these lines potentially allow us to further constrain the physical
conditions (e.g., density, temperature) within the headlight cloud.

\subsubsection{Observed line ratios in the headlight cloud}
\label{sss:ratiosHC}

\TabGaussianFits{} %
\TabLineRatHeadlight{} %
\TabLineRatGals{} %

The headlight cloud is detected in all molecular tracers except
CS(5$-$4). Moreover, it is the only source detected in these tracers in the
12-m ALMA observations. The locations of the emission peaks for the
different molecular lines match to within $<1$\arcsec\ for the CO lines and
to within $1\farcs8$ for the high density tracers. These higher density
tracers also have low signal-to-noise ratios (see
Fig.~\ref{fig:cloudtracerstpeak}). As a result it remains unclear whether
the offsets are significant. In the remainder of this article, we
will consider that the peaks for all lines are spatially coincident within
our current measurement accuracy. As a consequence, the left panel
  in Fig.~\ref{fig:lvg} shows the line profiles extracted at the position
  of the peak emission of each molecular line. To make a fair comparison,
we first smooth all images to match the resolution of the lowest resolution
dataset, $^{13}$CO(1$-$0), before extracting the line profile at the
emission peak (See Fig.~\ref{fig:cloudtracersIconv}).

We fit a single Gaussian profile to each line in Fig.~\ref{fig:lvg} and
report the results in Table~\ref{tab:fitprof}. We do not include the values
for HNC(1$-$0), since no reliable fit could be obtained. The centroid
velocities and linewidths of all tracers are in good agreement, although
HCN appears somewhat broader than the other lines. We have verified that
this is not due to the hyper-fine structure of the HCN(1$-$0) line, but it
may be an artifact of the limited velocity resolution and sensitivity of
the HCN data. Table~\ref{tab:linerat} lists the peak brightness and
integrated intensity line ratios for different combinations of the emission
lines. We use these ratios in Section~\ref{ss:lvg} to derive an estimate of the
density and temperature of the molecular gas in the headlight cloud.

\subsubsection{Comparison to literature line ratios}
\label{ss:line_ratios_literature}

\FigLineRat{} %
\FigRCOMass{} %

Here we compare the line ratios that we measure for the headlight cloud to
a compilation of line ratios in other Galactic and extragalactic
sources. We compare to nine nearby spiral galaxies that are part of the
EMPIRE survey \citep{bigiel16}, using the values determined by
\citet{cormier18} and \citet{jimenez-donaire19}. We also compare to line
ratios measured in Orion-B in the Milky Way \citep{sakamoto94,pety17} and 
30 Doradus in the LMC 
\citep{johansson98,pineda12,anderson14}. Table~\ref{tab:rat-gals} lists
these ratios, the associated spatial scales, and the references for the
measurements.

Figure~\ref{fig:rat-gals} shows the HCN/CO(1$-$0) line ratio in red,
HCO$^+$/HCN in green, $^{13}$CO/$^{12}$CO in yellow and \rco\ in blue. In
each case, we plot the line ratio in the comparison source
\textit{normalized to the value estimated for the headlight cloud, so that
  it indicates a line ratio identical to the headlight cloud}.  From the
top to bottom panel, we plot the data for entire galaxies, and for disk and
center regions separately. We do not show data for the inter-arm
  or diffuse extended regions. This is why the galactic ``average'' values
  can sometimes be lower than both the disk and center values. The line
ratios used in this plot and quoted in Table~\ref{tab:rat-gals} are derived
using different methods and on datasets with a range of resolution and
sensitivities, and the values should therefore be interpreted with some
caution. As described in Section~\ref{ss:other:tracers}, the values for
headlight cloud are determined using the peak brightness of a Gaussian fit
to line profiles extracted at a spatial scale of $\sim140$\,pc, whereas the
values for the EMPIRE galaxies are ratios calculated using the integrated
intensity of an average line profile from within a kpc-scale
aperture. A comparison between ratios measured at similar scales
  will only be possible when more high angular resolution data in nearby
  galaxies will become available. Using current data, the first striking
  point is that the ratios mostly exhibit the same order of magnitude
  independent of scale. This probably suggests that the emission of the
  different lines is co-spatial to first order, i.e., the lines arise from
  the same molecular gas phase. They are thus affected by beam dilution in
  similar ways. Detailed analysis of second order variations of the ratios
  may indicate other physical and chemical processes.

In extragalactic work, the ratio between HCN/CO ratio is often used as an
indicator of gas density. The ratio measured on 140\,pc scale in the
headlight cloud, HCN/CO = $0.025$ to 0.04 is in broad agreement with values
found in galaxy disks \citep[e.g.,][]{usero15,gallagher18}. It is higher
than the value measured on larger scales from the EMPIRE full-disk map of
NGC\,628 \citep{jimenez-donaire19}, for which the mean HCN/CO ratio is
$\sim0.015$. This can be seen from low normalized value, $\sim0.5$, of the
green symbol in NGC\,628 in Fig.~\ref{fig:rat-gals}, showing that the HCN/CO
in the headlight cloud is about two times higher than the galaxy-averaged
value. Assuming that the HCN/CO ratio highlights dense gas, the headlight
cloud appears to be denser than its surroundings, with more dense,
HCN-emitting gas, consistent with the compact structure at the core of the
cloud structure measured in CO.

The HCO$^+$/HCN ratio is more variable among the EMPIRE survey targets. The
HCO$^+$/HCN ratio in the headlight cloud is about twice the disk-averaged
value within NGC\,628 (0.55), and closer to the value estimated towards
NGC\,628's central region (0.89). The HCO$^{+}$/HCN ratio in the headlight
cloud is similar to the average ratio observed in Orion~B (1.1), but
significantly less than the ratio observed in the 30~Dor-10 molecular cloud
(5). We note that HCO$^{+}$/HCN ratios ranging between 3 and 10 have been
reported for several star-forming regions in the LMC \citep[including
e.g. N159W, N44, N105, and other clouds near 30~Dor,][]{seale12,
  anderson14} and the low metallicity Local Group dwarf IC~10
\citep{nishimura16,braine17,kepley18}. This may indicate a metallicity
dependence of this ratio, since HCO$^{+}$/HCN values close to unity are
observed in massive star-forming regions in our Galaxy
\citep[e.g. W51][]{watanabe17}, and values between 0.5$-$2 on larger scales
in starburst galaxies \citep[e.g.,][]{imanishi07,krips10,bemis19}.

The $^{12}$CO/$^{13}$CO line ratio in the headlight cloud is $7.2$, which
is comparable to the value in Orion~B \citep{pety17} and in the 30~Dor-10
molecular cloud. The EMPIRE measurements for NGC\,628 and other nearby
spiral galaxies are systematically higher, which we attribute to the lower
filling factor of $^{13}$CO emission within the EMPIRE resolution element
($27'' \sim 1$ to 2\,kpc).

The \rco\ ratio in the headlight cloud is $\sim0.7$, which is close to the
standard value for resolved observations in the Milky Way \citep[including
the Orion~B cloud, see e.g.,][]{sakamoto94, sakamoto97, yoda10}, and in
nearby galaxies \citep[e.g.,][]{eckart90,leroy09,leroy13,cormier18}. The
30~Dor~10 molecular cloud in the LMC \citep{pineda12} stands out as having
a \rco\ ratio close to unity, while the mean ratio for the kpc-scale
EMPIRE measurements in NGC\,628 and other nearby spiral galaxies is
slightly lower, $\sim0.6$, than in the headlight cloud. The EMPIRE \rco\
measurements show some variation between and within galaxies: in NGC\,5055
and NGC\,4321, the \rco\ ratio increases towards the galaxy centers, but
the opposite trend is seen for NGC\,628.

Since the kpc-scale of the EMPIRE measurements can obscure cloud-scale
variations of \rco\ within NGC\,628, we measured the mean \rco\ ratio
within a $3\farcs2 \times 3\farcs2$ box, centred on the positions of the
GMCs identified in NGC\,628 (see Section~\ref{ss:cprops}). To be
  consistent with the measurements of the headlight cloud in
  Section~\ref{sss:ratiosHC}, the line ratios were defined using the peak
brightness of CO(1$-$0) and CO(2$-$1) data cubes that had been smoothed to
common (round beam) resolution of $3\farcs2$. We restricted our measurement
to pixels where the signal-to-noise at the line peak was greater than 5 in
both tracers, and we rejected clouds where our analysis box contained less
than 10 valid pixels. We list the cloud-scale \rco\ measurements for
NGC\,628's ten most luminous clouds in Table~\ref{tab:clouds}, and we plot
all the cloud-scale measurements of \rco\ versus the cloud luminous mass in
Fig.~\ref{fig:r21-mass}. From this analysis, it is clear that kpc-scale
averages hide genuine local variations in \rco: we find a mean cloud-scale
value of \rco\ in NGC\,628 of 0.54, with a standard deviation of
$0.10$ in the cloud-scale measurements. We note here that the headlight
cloud has a relatively high \rco\ value relative to the rest of NGC\,628's
cloud population, but defer a detailed investigation of the physical origin
of \rco\ variations on sub-kpc scales to a future paper (Saito et al., in
prep).

\subsubsection{LVG modelling}
\label{ss:lvg}

Here we estimate the typical density and kinetic temperature of the
molecular gas of the headlight cloud using a Large Velocity Gradient (LVG)
analysis of some of the line ratios measured at the cloud's emission peak
(see Table~\ref{tab:linerat}).  We use
RADEX\footnote{\url{http://home.strw.leidenuniv.nl/~moldata/radex.html}}, a
public LVG radiative transfer code. We assume an expanding sphere geometry,
a velocity linewidth of 20\,\kms\ (this is the mean value of the observed
linewidth for all tracers, see column~4 of Table~\ref{tab:fitprof}), and a
background temperature of 2.73\,K. The H$_2$ column density is set to
$N_{\rm H_2}$ to 9$\times10^{21}$\,cm$^{-2}$, which is the value obtained
assuming a CO luminous mass of 2$\times10^7$~\msun\ and radius of 180\,pc
(see Sect.~\ref{ss:cprops:headlight}). We first compute a grid of LVG
models, which covers a kinetic gas temperature range of
$T_{\rm kin}=10-160$~K, and molecular gas density range of
$n_{\rm H_2}$=10$^2$~$-$~10$^7$~cm$^{-3}$. We fit the
$^{12}$CO(1$-$0)/$^{13}$CO(1$-$0) ratio as a good tracer of temperature
variations, HCN(1$-$0)/$^{13}$CO(1$-$0) as a tracer of gas density
variations and HCN(1$-$0)/$^{12}$CO(2$-$1) as a confirmation of the implied
temperature to density ratio.  The molecular abundances are fixed to their
typical Galactic values, {\it i.e.}, [$^{12}$CO]/[H$_2$] =
3$\times10^{-4}$, [$^{12}$CO]/[$^{13}$CO] = 70, and
[HCN]/[H$_2]= 5\times10^{-9}$~\citep{blake87,vandishoeck88}.

We do not attempt to fit the CO(2$-$1)/CO(1$-$0), and
HCO$^+$(1$-$0)/HCN(1$-$0) ratios as they are difficult to
interpret. Indeed, while the $^{12}$CO lines are optically thick, the
energy difference between the J=2$-$1 and 1$-$0 levels is only 17\,K. The
latter property makes the \rco\ ratio mostly sensitive to cold gas, while
the former property favors tracing diffuse gas at typical temperature of
80\,K. The HCO$^+$/HCN ratio is also difficult to interpret because it is
sensitive to the relative abundance of these species, which vary with
environment by an order of magnitude~\citep{goicoechea19}.

The right panel of Fig.~\ref{fig:lvg} shows the result of the fit of the
different line ratios as a set of color-coded thick lines for the possible
solutions and thin lines showing the 95\% confidence interval.  These
curves intersect at $n_{\rm H_2}\sim5\times 10^{4}$\,cm$^{-3}$ and
$T_{\rm kin}\sim20$\,K.  This is more than three orders of magnitude
greater than the average H$_2$ density estimated from the cloud mass and
diameter quoted above. Assuming that 10 to 100\% of the headlight cloud's
mass has this typical density and that the geometry is spherical, this
yields a typical radius ranging from 15 to 34\,pc. This gas has a typical
thermal pressure of $10^{6}$\,K\,cm$^{-3}$.

While this solution provides a good fit to the different line ratios, the
RADEX-predicted CO(2$-$1) brightness is about three times larger than what
is measured. We note this as a caveat to the inferred physical conditions,
and consider our results for the cloud density and pressure as
order-of-magnitude estimates. In view of the uncertainties in the beam
filling factor of the different emission lines, the appropriate
$X_\mathrm{CO}$ value, the molecule abundances, and the density
distribution adopted in the LVG model, it is difficult to justify more
sophisticated modelling with the current set of molecular line data.

\section{Current stellar feedback}
\label{ss:feedback}

In this section, we characterize the \ion{H}{ii} region associated with the
headlight cloud. In particular, we estimate the typical age and mass of the
associated young stellar population and we compute order-of-magnitude
pressures related to stellar feedback. More detailed modelling is beyond
the scope of this paper. It will be the subject of another forthcoming one.

\subsection{The headlight cloud's \ion{H}{ii} region}
\label{sec:hii}

The headlight cloud spatially coincides with the most luminous \ion{H}{ii}
region within the footprint of our MUSE observations. H$\alpha$ emission
from this region closely resembles the CO emission in both morphology and
kinematics, suggesting that the \ion{H}{ii} region is still embedded within
the cloud.  To within an accuracy of 1\arcsec{} (i.e., 47\,pc), the two
distributions peak at the same position. Figure~\ref{fig:mom1COHa} shows
that the velocity structure of the molecular and ionized gas is also very
similar.

As in CO, the H$\alpha$ emission associated with the headlight cloud is the
brightest region in the galaxy. This region also appears bright relative to
any simple extrapolation of the distribution of intensities from fainter
regions. We show this in Fig.~\ref{fig:moments:cdfa}. There, we plot
pixel-wise CDFs of the H$\alpha$ intensity $I_{\rm H\alpha}$ and
$W_{\rm CO}$ for the entire galaxy.

Figure~\ref{fig:moments:cdfa} shows that, like the CO emission, the
H$\alpha$ emission has a bump at high intensities. The sense of
this bump is that there are more bright pixels than one would
expect base on extrapolating from lower intensities. The bump is
specifically associated with the 45 brightest pixels in the H$\alpha$ map,
which we mark with a green line in Fig.~\ref{fig:moments:cdfa}. These
pixels are all associated with the headlight cloud. They belong to a region
2.3 square arcseconds in area that spatially correlates with the brightest
CO peak.

\subsection{Young stellar content}

\FigVelmapCOHa{} %
\FigCDFHa{} %

The extinction-corrected H$\alpha$ luminosity associated with the headlight
cloud~\citep{kreckel16,kreckel18} implies a SFR of 0.034\,\msun\,yr$^{-1}$
assuming a constant star formation history.  This is likely a lower limit
since the bulk of the star formation activity may still be
embedded. Moreover, the star formation history in the headlight cloud
region may be better described as an instantaneous burst (see below).

We use the above SFR estimate to assess the local molecular gas depletion
time $\tau_{\rm dep}^{\rm mol} \equiv M_{\rm mol}/{\rm SFR}$, finding
$\tau_{\rm dep}^{\rm mol} \sim 0.6\,$Gyr using $2.0 \times 10^7$~M$_\odot$
mass of the cloud. For comparison, we estimate
$\tau_{\rm dep}^{\rm mol} \sim 1.6$~Gyr for the $1$~kpc region around the
headlight cloud and $\sim 1.5$~Gyr for the whole area mapped by ALMA
\citep[both consistent with previous work
by][]{leroy08,bigiel08,leroy13,kreckel18}. The latter values represent a
large-scale equilibrium rate, while the former shorter value probably
represents a small spatial-scale snapshot that reflects the current
evolutionary state of the headlight
\citep[see][]{schruba10,kruijssen14,kruijssen18}.

\TabHalphaProp{} %

We can obtain a sharper picture of the recent star formation embedded
within the headlight cloud by comparing observed properties to Starburst99
(S99) models assuming an instantaneous burst \citep{leitherer99}. In this
case, our best constraint on the age comes from the equivalent width of the
H$\alpha$ line, and our best constraint on the mass comes from the number
of ionizing photons produced, as indicated by the total H$\alpha$ emission.
The observations were compared to models using a solar metallicity, an
instantaneous burst of star formation of 10$^6$~\msun, and a Kroupa initial
mass function (IMF).

The age is estimated to be 4\,Myr from the measured H$\alpha$ equivalent
width (EW) of $517$~\AA. This is an upper limit since S99 does not yield
accurate age measurements when the region is younger than $4$~Myr. The MUSE
spectrum of this \ion{H}{ii} region shows the presence of the \ion{C}{IV}
line at $\lambda$5801-12\AA. This line is a specific feature of Wolf-Rayet
stars, which are present in stellar populations aged between 2 and
6~Myr. Combining both results, the age of the young stellar population is
constrained to 2${-}$4 Myr.

The mass in young stars is estimated from the number of ionizing photons,
$N_{\rm ion}$ they produce, itself estimated from the integrated
extinction-corrected H$\alpha$ luminosity of the headlight cloud's
\ion{H}{ii} region $L($H$\alpha)= 6.3 \times 10^{39}$~erg~s$^{-1}$. Using
Eq.~5 from \citet{calzetti10}, this luminosity yields
$N_{\rm ion}=4.6\times 10^{51}\,$photon\,s$^{-1}$.  The S99 single stellar
population models assuming this value of $N_{\rm ion}$ yield a stellar mass
as large as $\sim 3 \times 10^5$~\msun{} and a bolometric luminosity of
$L_{\rm cl}=4\times 10^8$~L$_{\odot}.$ The measured and estimated
properties of the young massive stellar population embedded in the
headlight cloud are summarized in Table~\ref{tab:halpha-prop}.

\subsection{Stellar Feedback}

The spatial coincidence of a massive young stellar population and massive
GMC indicates that these stars have not yet dispersed their parent
molecular cloud. In this section, we investigate the potential feedback
mechanisms that may lead to the disruption of the headlight cloud, adopting
the properties for the cloud and the young stellar population listed in
Tables~\ref{tab:clouds} and \ref{tab:halpha-prop}, respectively.

The population of Wolf-Rayet stars will produce strong winds that will push
away the low density ionized gas of the \ion{H}{ii} region. In the simple
spherically symmetric picture, this will results in a thin or thick shell
of ionized gas surrounding an almost empty central region. The density in
the shell will be higher than the mean density of the sphere, so that the
ram pressure approximately balances the thermal pressure.

To examine the balance of forces in more detail, we start by assuming that
the hot gas associated with the wind has been lost from the young stars,
either because it has already radiatively cooled or because it has escaped
through low density gaps in the shell \citep[see e.g.][]{rogers13}. This
assumption is consistent with the fact that we do not find any signature of
compact X-ray emission associated with the headlight cloud in Chandra and
XMM-Newton observations \citep{liu05b,owen09}. It is also consistent with
detailed models of the effects of stellar winds on molecular clouds, which
typically find that the hot gas is lost within only $\sim 1$~Myr from the
onset of the wind \citep[see e.g.][]{rahner17}. The mechanical luminosity
of the wind predicted by the S99 models is
$L_{\rm mech}=4.9\times 10^6$~L$_{\odot}$. This can also be written in
terms of the total stellar mass loss rate ($\dot{M}$) through
\begin{equation}
  L_\emr{mech} = \frac{1}{2} v_\emr{wind}^{2} \dot{M},
\end{equation}
where $v_{\rm wind}$ is the characteristic wind velocity. In the present
case, the wind luminosity is dominated by the Wolf Rayet stars and so the
wind velocity will typically be very large, i.e.,
$v_{\rm wind} \sim 2\times10^3\,$km\,s$^{-1}$. This yields a mass loss rate
$\dot{M} \sim 9.4 \times 10^{23}\,$g\,s$^{-1}$. The ram pressure at radius
$R$ is
\begin{equation}
  P_{\rm ram,wind} = \frac{v_{\rm wind} \dot{M}}{4 \pi R^2}.
\end{equation}
Assuming that $R$ is the radius of the \ion{H}{ii} region, we find 
$P_{\rm ram,wind} \sim 5.7 \times 10^5\,$K\,cm$^{-3}$. This value depends
on our assumed wind velocity, since
$P_{\rm ram,wind} \propto v_{\rm wind}^{-1}$, but does not change by more
than a factor of two for wind velocities in the range consistent with Wolf
Rayet stars.

If some of the massive young stars have already exploded as
supernovae (SN), these will also contribute to the total ram
pressure. The importance of this contribution depends on the age of the
young stellar population and the assumed star formation history. For our
assumption of an instantaneous burst, the first supernovae occur at a
time $t \sim 3.6$~Myr \citep{leitherer99}.  The headlight cloud appears
as a compact source in the non-thermal emission map at 3.1\,GHz in
  \citet{mulcahy17}, thus likely tracing synchrotron emission from
  supernovae.  We can estimate the contribution that the associated
  supernovae make to the ram pressure from
  \begin{equation}
    P_{\rm ram,SN} = \frac{v_{\rm SN} \dot{M}_{\rm SN}}{4 \pi R^{2}}.
  \end{equation}
  Here $\dot{M}_{\rm SN}$ is the supernova mass loss rate, which we take
  from the S99 model, and $v_{\rm SN}$ is the terminal velocity of the
  supernova ejecta, which we take to be $10^{4}$~\kms,
  following \citet{rahner17}. Note that this estimate is conservative, in
  that it neglects any contribution made by the pressure of the hot gas
  associated with the supernovae, consistent with our treatment of the
  stellar winds. For a young stellar population age of 4\,Myr, this yields
  a ram pressure contribution
  $P_{\rm ram, SN} \sim 3.0 \times 10^{5} \, {\rm K \, cm^{-3}}$, i.e.,
  roughly half of the contribution coming from stellar winds. We recover
  similar results for other assumed ages in the range
  $3.6 < t < 6.0$\,Myr.

At the radius of the \ion{H}{ii} region, the wind ram pressure balances the
thermal pressure due to the ionized photons \citep{pellegrini07}, provided
that other sources of support (e.g.\ magnetic fields) are unimportant. If
we assume that thermal pressure is the main source of support for the
ionized shell, and that the gas has a typical \ion{H}{ii} region
temperature of $10^4\,$K, then it follows that the ionized gas must have a
typical electron density of $n_\emr{e} \sim 30\,$cm$^{-3}$.  If we assume
that this density is achieved in a shell with radius $R_\ion{H}{ii}$, we
can also solve for the shell thickness $(dR)$ required to have ionization
balance
\begin{equation}
  N_\emr{ion} = \frac{4 \pi}{3} \alpha_B n_\emr{e} R_\ion{H}{ii}^2 dR,
\end{equation}
where $\alpha_B$ is the case B recombination coefficient. This gives
$dR \sim 7\,$pc. Note that this is technically the thickness of the ionized
portion of the shell: if the expansion of the shell has also swept up a
substantial amount of neutral/molecular gas, then the full shell will be
even thicker. Finally, we can calculate the mass of the ionized gas:
$M_\emr{ion} \sim 6 \times 10^5\,\msun.$ The ionized gas thus does not
contribute much to the overall mass budget. For comparison,
\citet{galvan13} find that in W49 (one of the most massive GMCs in the
Milky Way, albeit a factor 10 less massive than the headlight cloud) only
$\sim1\%$ of the total gas mass is in the ionized component, comparable to
the value of a few percent we infer here for the headlight cloud.

The radiation pressure is estimated following Eq.~1 in
\citet{herrera17}. For a radius defined by the H$\alpha$ emission, and
assuming no trapping of the IR photons within the shell, the radiation
pressure is $P_{\rm r0ad}=1.5\times 10^5$~K~cm$^{-3}$. The turbulent
pressure in the molecular gas, $P_{\rm turb}=\rho \sigma_{v}^2$, is
estimated to be $P_{\rm turb}=1.9 \times 10^5~{\rm K~cm^{-3}}$.

The magnetic pressure cannot be accurately computed, as we have no
  measurement of the magnetic field strength on the scale of the headlight
  cloud. If we adopt a value of $10~\mu$G, based on the kpc-scale
  measurements of \citet{mulcahy17}, the resulting magnetic pressure is
  $P_{\rm mag} = 2.9 \times 10^{4} \, {\rm K \, cm^{-3}}$, implying that
  the field would not play a significant role in regulating the expansion
  of the \ion{H}{ii} region \citep{vanmarle15}. It is possible that the
  cloud-scale magnetic field strength is larger than this, implying a
  larger magnetic pressure, but we do not expect $P_{\rm mag}$ to exceed
  the turbulent pressure unless the turbulence in the cloud is
  substantially sub-Alfv\'enic. Since it would be difficult to explain the
  large amounts of recent star formation if the cloud were strongly
  magnetically-dominated, we conclude that $P_{\rm mag}$ is unlikely to
  greatly exceed $P_{\rm turb}$.

The gravitational pressure exerted by the young stellar body and the
molecular cloud itself is $P_{\rm grav} = 3.6\times 10^5$~K~cm$^{-3}$,
following Eq.~5 in \citet{herrera17}, assuming a size defined by the CO
emission. In this computation, we neglected the contribution of the ionized
component as discussed above. Hence, the combined outward
  pressures, $P_{\rm turb}$, $P_{\rm ram}$ (winds and supernovae), and
  $P_{\rm rad}$, overcome the gravitational and magnetic pressures by a
  factor 2-4.

\subsection{Comparison with 30\,Dor in the LMC}

We note that the properties of the \ion{H}{ii} region in the headlight
cloud are similar to N157A \citep{henize56}, the \ion{H}{ii} region
associated with the 30\,Dor star-forming region, which is the largest and
most active star-forming region in the Local Group. N157A has an ionizing
photon rate of $N_{\rm ion}= 4.5\times10^{51}$\,photon\,s$^{-1}$ and a size
of $\sim$200\,pc \citep{walborn91}; the \ion{H}{ii} region in the headlight
cloud has $N_{\rm ion}= 4.6\times10^{51}$\,photon\,s$^{-1}$ and a radius of
$\sim$140\,pc (see Table~\ref{tab:halpha-prop}).  The age and mass of the
embedded young stars in the headlight cloud are likewise comparable to
30\,Dor's central star cluster (R136), which is $\sim$1-3~Myr old and has a
stellar mass of $4.5 \times 10^5\,\msun$ \citep{bosch09}. The molecular gas
properties of these regions are quite different, however. The integrated
$^{12}$CO(1$-$0) luminosity that is observed to be spatially projected onto
N157A is $\sim 2.5 \times 10^5$\,K\,km\,s$^{-1}$\,pc$^{2}$, corresponding
to only $\sim10^{6}\,\msun$ of molecular gas for a Galactic CO-to-H$_{2}$
conversion factor \citep{wong11,johansson98}. The young stars in 30~Dor
thus appear to have cleared out most of the molecular gas associated with
the \ion{H}{ii} region, contrary to the headlight cloud, for which we
estimate a molecular mass of $\geq10^{7}\,\msun$.

\section{The Galactic Environment of the Headlight Cloud}
\label{ss:galaxy_environment}

\FigRgalProps{} %

Both the properties and location of the headlight cloud make it remarkable,
as massive star-forming regions usually tend to appear at bar ends or
galaxy centers. In this section, we examine the environment of the
headlight cloud. First, we compare the properties of the region around the
cloud to the rest of the galaxy. Then we consider how the cloud's location
relates to the dynamical structure of the galaxy.

\subsection{Trends with Galactocentric Radius}
\label{ss:ISMenvironment}

In Fig.~\ref{fig:rgal_props}, we show the kpc-scale structure of NGC\,628,
highlighting the location of the headlight cloud. We plot the gas and
stellar structure in NGC\,628 as a function of galactocentric radius. Each
point in Fig.~\ref{fig:rgal_props} shows mean properties of NGC\,628
estimated within a 1\,kpc aperture~\citep[see][for
details]{leroy16,sun18,utomo18}. The apertures cover the galaxy. The
measurements are taken from Sun et al. (in prep.).

The measurements shown in Fig.~\ref{fig:rgal_props} are derived from
multi-wavelength archival data for NGC\,628. In addition to the PHANGS-ALMA
CO (2$-$1) maps, we use THINGS HI 21\,cm moment-0 map~\citep{walter08} to
trace atomic gas distribution, S$^4$G dust-subtracted $3.6\,\mu m$
image~\citep{sheth10,querejeta15} to trace stellar mass distribution, and
GALEX-NUV and WISE band~3 images \citep[presented in ][]{leroy19} to trace distribution of star formation activities. We translate these
observations into physical units (e.g., $\msun\,{\rm pc^{-2}}$) following
standard techniques, as presented in the aforementioned publications.  We
plot the stellar surface density, atomic gas surface density, molecular gas
surface density, and the estimated dynamical equilibrium pressure, all as a
function of radius. We mark the location of the region containing the
headlight cloud in red and show the environments containing the next
brightest clouds (i.e., the other nine clouds in Table \ref{tab:clouds}) in
blue.

The top left panel in Fig.~\ref{fig:rgal_props} shows stellar surface
density as a function of radius. The stellar mass in NGC\,628 is dominated
by a relatively smooth exponential disk. Here the regions hosting bright
clouds appear unremarkable. Many lie far from the galaxy center
($\sim 3{-}5$~kpc) and at comparatively low stellar surface density. Both
the large radius and low stellar surface density may be surprising. As
emphasized above, dense gas and high surface density gas tend to be more
common in galaxy centers \citep[e.g, see][and references
therein]{gallagher18,sun18}. Meanwhile, the stellar surface density plays a
key role in setting the ISM equilibrium pressure
\citep[e.g.,][]{elmegreen89,ostriker10}. ISM pressure has been found to
correlate with the prevalence and properties of molecular gas
\citep[e.g.,][]{wong02,blitz06,leroy08,hughes13b}. These clouds lie at
large radius and low stellar surface density, and so we might expect the
gas to be mostly atomic and full of clouds with modest internal pressures.

The top right and bottom left panels show that the atomic gas and molecular
gas surface densities in NGC\,628 become comparable beyond
$\sim4{-}5$\,kpc, while the inner part of the galaxy is more molecule
rich. While bright clouds are found at these radii, the bottom right panels
shows that they tend to be found in environments where the gas is
predominantly molecular. Outside the galaxy center, the molecular gas
surface density shows large spread at fixed radius, often $\sim 0.5$~dex (a
factor of three) or more.  This reflects that the strong spiral structure
dominates the morphology of the galaxy, leading to a high degree of
azimuthal structure. The bright clouds all appear among the highest surface
density points at their radii. Moreover, the regions with bright clouds at
$r_{\rm gal} \sim 3{-}4$~kpc have molecular gas surface densities
comparable to the galaxy center. This did not have to be the case, making
regions with bright clouds remarkable in NGC\,628.

The last panel combines the information from the first three panels
following \citet{ostriker10} and \citet{elmegreen89} to estimate the mean
dynamical equilibrium pressure. The formulae and detailed calculation is
presented in Sun et al. (in prep.) but the version plotted here follows
\citet{gallagher18} closely. See that paper for more
details. Qualitatively, this quantity represents the mean pressure needed
to support the weight of the ISM due to both self-gravity and stars. As
mentioned above, it has been shown to correlate with both molecular content
and the properties of molecular clouds.

The bright clouds, including the headlight cloud, all lie at the high end
of the pressures found in NGC\,628. This is true despite the comparatively
low stellar surface density, indicating that self gravity of the gas plays
a large role here. This, along with the significant azimuthal scatter also
visible in the bottom right panel, again emphasizes the important role of
galactic dynamics in creating these clouds.

Figure~\ref{fig:rgal_props} paints a picture of a relatively quiescent
galaxy in which the location of massive clouds is driven by the spiral
structure. NGC\,628 lacks a bulge, stellar bar or other feature to drive
nuclear gas concentrations. Its overall surface density of gas and stars is
modest. As a result, the concentration of gas by spiral arms creates some
of the highest pressure regions in the galaxy, with the pressure driven by
gas self-gravity.  The bright clouds identified by CPROPS, including the
headlight cloud, preferentially fall along these arms.

Given the apparent central role of dynamics in concentrating molecular gas
to create these clouds, we turn our attention to this topic in the next
section.

\FigCorotation{}%

\subsection{Relation to the spiral arms and radial inflows}
\label{sec:dynqualities}

Here we consider the particular properties and star formation activity of
NGC\,628's headlight cloud in relation to the large-scale dynamical
structure of the disk. Figure~\ref{fig:m74corotation} shows the spatial
distribution of GMCs in the Rosolowsky et al. (in prep.) catalog, using
colour to represent the CO luminous mass, velocity dispersion, virial
parameter and molecular gas surface density of the clouds.  The size of the
symbol corresponds to the cloud size. The shaded light yellow circles
represent corotation radii of 3.2\,kpc (red circles) and 5.1~kpc (green
circles) \citep[see text below, ][]{cepa90,fathi07}, with an uncertainty of
15\%.

It is evident from Fig.~\ref{fig:m74corotation} that, as suggested in the
previous section, large, massive clouds in NGC\,628, including the
headlight cloud, are located preferentially in the two prominent, tightly
wrapped, spatially extended spiral arms.  This suggests that the pattern of
gas flow along and through the spiral arms may be important for determining
the growth, longevity and properties of these clouds.  The gas response to
the spiral pattern has already been studied in detail by
\citet{fathi07}. They applied the Tremaine-Weinberg method to the
ionized gas kinematics to measure a pattern speed $\Omega_{\rm p}$ that is
consistent with the corotation radius at 5.1\,kpc identified earlier by
\citet{cepa90}. This cororation radius lies near the edge of our map (green
circles in Fig.~\ref{fig:m74corotation}).

In the presence of a single spiral pattern, with a single corotation
radius, all the clouds within our field-of-view, including the headlight,
would interact with the spiral pattern on a timescale of
$\tau_{\rm sp}=\pi/(\Omega-\Omega_{\rm p})$. This yields
$\tau_{\rm sp}=140$\,Myr at the location of the headlight cloud.  However,
from the detailed kinematic analysis of \citet{fathi07}, we infer a second,
inner corotation radius at $R\approx65$\arcsec$=3.2$\,kpc, very near the
position of the headlight cloud.  In this case, the headlight would remain
essentially motionless with respect to the spiral pattern over its entire
life, as discussed further below.

Following \citet{wong04}, the second corotation can be inferred from the
pattern in the coefficients of the harmonic decomposition of the
line-of-sight velocity field measured by \citet{fathi07}.  We have sought
to confirm this inner corotation radius using a new measurement of
gravitational torques across NGC\,628. We follow the method of
\citet{querejeta16} and \citet{garciaburillo05}, which requires a map of
the 2D stellar mass distribution to infer the underlying disk volume
density and gravitational potential $\Phi$, and a map of the gas
distribution to provide a record of the time-averaged position of the gas
relative to the potential. We use the 3.6$\mu$m contaminant-cleaned map
produced by S$^4$G Pipeline 5 \citep{querejeta15} for the former, adopting
a global $M/L=0.6$ assuming a Chabrier IMF \citep{meidt14}. We use the ALMA
map shown in the right panel of Fig.~\ref{fig:moments:full} to trace the
gas mass distribution. Following \citet{garciaburillo05}, we measure the
azimuthally-averaged radial torque profile as
\begin{equation}
  \tau(R)=\frac{\int_\theta [N(x,y) (xF_y-yF_x)]}{\int_\theta N(x,y)},
\end{equation}
where $R$ and $x,y$ are the radial and Cartesian coordinates, $N(x,y)$ is
the gas column density, and the gravitational forces are computed from the
reconstructed potential $\Phi$ with $F_{x,y}=\nabla_{x,y}\Phi$.  We then
measure the differential mass flow rate as
\begin{equation}
  \frac{d^2M(r)}{dRdt}=\tau(R)\frac{1}{L}\bigg\vert_\theta 2\pi R N(x,y)\vert_\theta,
\end{equation}
assuming that the azimuthal average of the angular momentum is
$L_\theta$=$RV_{\rm rot}$.  The resulting profile of inflow driven by
gravitational torques is shown in the
Fig.~\ref{fig:dynstructure}\footnote{The largest uncertainty on the value
  at any given location is the stellar mass to light ratio used to convert
  the 3.6$\mu$m contaminant-cleaned map into a stellar mass map. Although
  spatial variations in the $M/L$ can lead to variations in the measured
  torques and gas inflow rates, they are not expected to greatly alter the
  pattern of sign changes in the torque, which we use to infer the
  dynamical structure of the spiral pattern. For a detailed assessment of
  the measurement uncertainty associated with this method of determining
  the torque profile, see \citet{querejeta16}.}. Overall, the torque
experienced by the gas in NGC\,628 undergoes several sign changes,
including a crossing from negative to positive near
$R=65$\arcsec$=3.2$\,kpc, which marks the location of our inner corotation
radius, CoR$_{\rm in}$, and corresponds to the galactocentric radius of the
headlight cloud.

\FigDynStructure{} %

The position of the headlight cloud very near the location of the inner
corotation radius may have important implications for its longevity and
ability to grow to its present mass, and thus the vigorous star formation
occurring there.  As indicated in the top panel of
Fig.~\ref{fig:dynstructure}, gravitational torques and (differential) gas
flows are zero at corotation.  Thus, compared to other locations along the
gaseous spiral arms, corotation is a relatively sheltered environment: a
cloud at corotation radius is less susceptible to destructive dynamical
forces such as shear, and rotates together with the spiral pattern during
its entire evolution (i.e.  $t_{\rm sp}\gg t_{\rm orb}$). Another factor
that may promote the growth of the headlight cloud is that NGC\,628's inner
corotation radius overlaps with one of the inner ultra-harmonic resonances
(UHR) of the outer spiral pattern UHR$_{\rm out}$, as illustrated in the
bottom panel Fig.~\ref{fig:dynstructure}. In this case, gas sitting in the
outer disk that is torqued by the outer spiral would be driven radially
inward from the outer corotation radius, CoR$_{\rm out}$, towards the
UHR$_{\rm out}$, providing a continual supply of gas to this region around
CoR$_{\rm in}$. Following \citet{querejeta16}, we use the profile of
gravitational torques to measure the radial flow across the disk of
NGC\,628, which we show in the middle panel of Fig.~\ref{fig:dynstructure}.
From the measured $d^2M/dt\,dR$, we estimate that the net mass flow between
the inner and outer corotation radius is indeed radially inward, at a net
rate $0.07\pm0.04$~\msun~yr$^{-1}$.  The inflow rate is a lower limit since
it does not take into account a contribution from viscous torques
\citep[which are expected to be dominated by gravity
torques;][]{combes90,barnes96}. We develop these ideas as a possible
explanation for the mass growth of the headlight cloud in
Section~\ref{sec:dyndiscussion}.

\section{Discussion}
\label{ss:discussion}

We have identified the most luminous molecular cloud in NGC\,628, the
headlight cloud, which is $\sim 2.4$ times brighter in CO(2$-$1) than any
other molecular cloud in NGC\,628. It is the most massive cloud
$(M_{\rm lum}=2.0\times10^7\,\msun)$ in NGC\,628. The cloud is located in a
spiral arm at a galactocentric radius of $\sim3.2$\,kpc, which is
coincident with a spiral corotation radius. The cloud itself hosts the most
luminous \ion{H}{ii} region in NGC\,628\ detected by MUSE, which is
associated with a young ($2-4$~Myr) massive ($3\times10^5$~\msun) stellar
population.

\subsection{Overluminous CO emission in the headlight cloud}
\label{sec:overluminous}

The central CO(2$-$1) integrated intensity and peak brightness temperature
of the headlight cloud are high compared to the other GMCs in NGC\,628,
including other massive clouds. While a large integrated intensity could be
due to a large cloud linewidth, Fig.~\ref{fig:moments:full} shows that the
headlight cloud's peak temperature is also $\sim2$ times greater than other
clouds in NGC\,628.

More quantitatively, at the $\sim40$\,pc resolution of our
  CO(2$-$1) data, the average peak brightness temperature of the CO(2$-$1)
  emission from clouds with $M_{\rm vir} \geq 10^{6}$\,\msun\ in NGC\,628\
  is $T_\emr{mb}=1.3$\,K. The headlight cloud peak temperature is
  6.7\,K. Making a local thermodynamic equilibrium simulation of the radiative transfer, which assumes
  that the headlight cloud is composed of ``standard'' molecular gas (i.e.,
  with the characteristic kinetic temperature of 10\,K,
  \citealt[e.g.][]{krumholz11}) filling the full cloud volume, the
  predicted peak temperature would only be 5.5\,K, i.e., less than the
  measured one. Alternatively, a typical kinetic temperature of 50\,K would
  be required so that the headlight cloud has the same beam filling factor
  of a typical GMC of $10^6$\,\msun{} in NGC\,628. Hence, although our
  measurements do not exclude that there may be an additional increase in
  the filling factor of the CO(2$-$1) emission in the headlight cloud
  region, we favor an increase of the excitation temperature above 10\,K to
  explain the headlight cloud's peak brightness because of the spatial
  coincidence with the HII region heating source.

Another clue that heating is important comes from the \rco\ ratio, which we
measured on 140\,pc scales for the cloud population in NGC\,628\ in
Fig.~\ref{fig:r21-mass}. This ratio is $\sim$0.7 in the headlight cloud,
higher than the average ratio for the other bright clouds in
Table~\ref{tab:clouds} (0.6), and higher than the average ratio for the
overall cloud population (0.54). One possibility is that intense external
heating could cause a positive temperature gradient across the photospheres
of CO-emitting clumps within the headlight cloud. Due to its larger
opacity, the CO(2$-$1) transition might then preferentially sample a warmer
clump layer than the CO(1$-$0) transition, raising the \rco\ ratio. We
defer a more detailed investigation of the physical factors linked to
variations in the \rco\ ratio in the disks of PHANGS galaxies to a future
paper.

Almost all of the luminous GMCs in NGC\,628\ listed in
Table~\ref{tab:clouds}, including the headlight cloud, have low virial
parameters (see Fig.~\ref{fig:larson}) as determined from the ratio between
their CO luminosity and the virial mass estimates. If the CO(2$-$1) line is
over-luminous due to excitation effects, then the true mass of the
headlight cloud is smaller than the luminous mass estimate of
$2.0\times10^{7}$~\msun. The low virial parameter and bright CO(2$-$1)
emission in the headlight cloud is thus consistent with the suggestion
by~\citet{pety17} that mass estimates based on the CO luminosity may be
biased when GMCs are closely associated with H$\alpha$ regions, as observed
for the Milky Way Orion~B cloud. If so, then clouds associated with
\ion{H}{ii} regions should show lower apparent virial parameters than
otherwise identical clouds without nearby heating sources. We intend to
test this hypothesis in a forthcoming paper using the full sample of PHANGS
targets with cloud-scale imaging by ALMA and MUSE.

\subsection{Feedback-limited lifetimes of clouds \& prolonged survival of
  the most massive objects}

The intense feedback from the newly-formed embedded stellar population in
the headlight cloud is expected to lead to its eventual destruction. We
appear to be observing the headlight cloud just before being destroyed by
the feedback from the present generation of stars.  In the simulations
of~\citet[][see also \citealt{dale17}]{kim18}, massive clouds with higher
escape velocities can withstand destruction via photoionization and
radiative pressure longer than their lower mass counterparts.  For a cloud
with the headlight's properties, $t_\emr{dest}\sim 10\,$Myr \citep{kim18},
which is longer than the age of the headlight's young stellar population.
If clouds with larger masses are able to survive for longer in the presence
of feedback, then the overluminous state of the headlight cloud is linked
not only to its star formation activity, but also to its large mass.

\subsection{The influence of galaxy dynamics on cloud growth}
\label{sec:dyndiscussion}

If the mass of the headlight cloud is key to its longer survival time in
the presence of feedback from star formation, the next question is what
mechanism(s) allowed it to grow to its extraordinary mass.  The dynamical
environment of the headlight offers some clues.

Massive cloud formation and incipient starbursts are often linked to
extreme conditions, such as in galaxy centers, at the ends of bars and in
mergers
\citep[e.g.][]{kenneylord91,elm94,boeker08,kenn98,kormkenn,elm2011,beuther},
where elevated pressures can confine highly turbulent star-forming gas
\citep[e.g.][]{herrera17,johnson15,leroy18}.  The normal disk environment
may also provide a less extreme avenue for massive cloud growth (Meidt \&
Kruijssen, in prep.).  As suggested earlier in
Sect.~\ref{sec:dynqualities}, the spiral corotation radius where the
headlight is positioned may furnish a favorable environment for the growth
of particularly massive clouds.

At the corotation radius of a spiral pattern, the pattern moves at the same
velocity as material in the galaxy disk. An individual cloud at the
intersection of a spiral arm with a corotation radius will thus reside
within the spiral pattern for its entire life.  Here, gas experiences no
gravitational torques and shear is locally reduced, promoting the longevity
(and growth) of molecular clouds situated in this zone in the absence of
feedback. In Section~\ref{sec:dynqualities}, we suggested that the
headlight cloud's position at the inner corotation radius may further
support the mass growth of the cloud via gas flows driven into the region
by gravitational torques.  Normally, an isolated corotation radius is
expected to be relatively devoid of gas \citep{belmegreen96}, due to the
inflow (outflow) of material radially inward (outward) from this point
\citep[e.g.][]{garciaburillo94,meidt13,querejeta15}. However, NGC 628's
inner corotation radius appears to coincide with an inner dynamical
resonance of the outer, independently rotating, spiral pattern (see bottom
panel of Fig.~\ref{fig:dynstructure}).  This type of scenario has been
identified in simulations \citep{tagger87,rautiainen99} and observations
\citep{delmegreen02,meidt09,font19}. In this case, the radial inflow of gas
possible from the outer corotation radius inward towards the inner
corotation radius provides a potential mass reservoir for the cloud’s
exceptional growth (Meidt \& Kruijssen, in prep), raising its mass above
the fiducial level set by competition between self-gravity, shear and
feedback \citep[e.g.,][]{reina17}.

We can use the mass inflow rate measured in Section~\ref{sec:dynqualities}
to estimate the timescale for the growth of the headlight cloud.  On one
hand, we can place a lower bound on this timescale, assuming that all of
the material inflowing from the outer corotation radius is accreted onto
it.  The measured inflow rate in this case implies that the cloud grew to
its present mass in time $t_{\rm grow}\sim 215\pm100\,$Myr, which is
roughly two orbital periods at the cloud’s location.  On the other hand,
considering that the inflow of gas onto the cloud is happening on multiple
scales, this estimate can be viewed as an upper limit on the true growth
timescale.  Indeed, the mechanisms by which that matter accretes onto the
cloud \citep[e.g. turbulent or gravitationally-driven
accretion,][]{ibanez17} may incorporate local material that does not
originate in the outer disk, allowing condensation to proceed much more
rapidly.  Viscous torques would also imply a shorter growth timescale, by
increasing the mass inflow rate to the region.

\section{Conclusions}
\label{ss:conclusions}

We have presented a detailed study of the molecular line emission and star
formation activity associated with the headlight cloud. This cloud is the
most massive molecular cloud in NGC\,628, and it hosts an exceptionally
bright \ion{H}{ii} region identified by MUSE \citep{kreckel18}. Our
objective was to characterize the properties of this molecular cloud in the
context of the internal (feedback) and external (dynamical) factors that
may influence its formation and evolution.

Our main science results are:
\begin{itemize}
\item The headlight is 2.4 times brighter than any other cloud in
  CO(2$-$1), and it is the only source with HCO$^{+}$(1$-$0), HCN(1$-$0)
  and HNC(1$-$0) emission detected by the ALMA 12-m array. The cloud is
  associated with the most luminous \ion{H}{ii} region in NGC\,628.
\item The headlight cloud is the most massive
  ($M_{\rm lum}= 2.0\times 10^7$\,\msun) cloud in NGC\,628. Its virial
  parameter is low, $\alpha_{\rm vir}=0.5$. We argue that the CO emission
  is over-luminous based on its high CO(2$-$1) peak temperature and its
  spatial coincidence with a bright peak in the H$\alpha$ emission. This
  implies that its true molecular mass would be located between its virial
  mass of $9\times 10^6$\,\msun{} and its luminous mass, $M_{\rm lum}$.
\item According to a simple LVG analysis of the multi-line ALMA data, the
  typical density and temperature in the headlight cloud is
  $n_{\rm H_2}\sim5\times 10^{4}$\,cm$^{-3}$ and $T_{\rm kin}\sim20$\,K.
\item From the presence of Wolf-Rayet features in the MUSE spectrum and
  modelling of the H$\alpha$ emission using Starburst99, we infer the
  presence of a young (2$-$4\,Myr), massive ($3\times 10^5$\,\msun) stellar
  population that is still embedded inside the headlight cloud.
\item We propose that feedback from the vigorous star formation happening
  within the headlight is responsible for its apparent overluminous state,
  captured just prior to the destruction of the cloud. This is possible
  given the longer survival times expected for the most massive objects
  \citep[e.g.,][]{kim18}.
\item We propose that the extreme headlight mass within the NGC\,628 disk
  is related to the large scale dynamics of the galaxy.  In particular, the
  cloud's location at the intersection of a spiral arm and the inner
  corotation radius implies not only a special location against the
  disruptive effects of galactic shear, but also a sustained inflow of gas
  favorable to enhanced mass growth. This scenario, in which clouds can
  grow to exceptional masses outside more prototypical `extreme'
  environments, will be discussed in more detail in a future paper by Meidt
  \& Kruijssen (in prep).
\end{itemize}
The exceptional star-forming activity and physical properties of the
headlight cloud call for deeper, higher angular resolution ALMA
observations of tracers of the molecular gas density distribution within
the cloud.

In addition to our scientific results, we developed our own CASA data
reduction procedures for the ALMA Total Power data in order to produce the
position-position-velocity cube combining the ALMA 12m, 7m, and total power
observations of NGC\,628. These procedures, which are used in the reduction
of all the Total Power data obtained by observational programmes led by the
PHANGS consortium (e.g., Leroy et al., in prep.), are described in
Appendix~\ref{app:tp}. As part of our PHANGS-ALMA Large Programme
commitments, the procedures have been transferred to the Joint ALMA
Observatory so that they can be adapted for general use. Scripts
implementing our procedures are also available on GitHub at
\url{https://sites.google.com/view/phangs/home/data}.

\begin{acknowledgements}
  We thank the anonymous referee for a prompt and helpful report that 
  strengthened our manuscript.
  This paper makes use of the following ALMA data:
  ADS/JAO.ALMA\#2012.1.00650.S and 2013.1.00532.S. ALMA is a partnership of
  ESO (representing its member states), NSF (USA) and NINS (Japan),
  together with NRC (Canada), NSC and ASIAA (Taiwan), and KASI (Republic of
  Korea), in cooperation with the Republic of Chile. The Joint ALMA
  Observatory is operated by ESO, AUI/NRAO and NAOJ.  The National Radio
  Astronomy Observatory is a facility of the National Science Foundation
  operated under cooperative agreement by Associated Universities, Inc. We
  thank J. R. Goicochea for useful discussions. CH, AH and JP acknowledge
  support from the Programme National ``Physique et Chimie du Milieu
  Interstellaire'' (PCMI) of CNRS/INSU with INC/INP co-funded by CEA and
  CNES, and from the Programme National Cosmology and Galaxies (PNCG) of
  CNRS/INSU with INP and IN2P3, co-funded by CEA and CNES. The work of AKL,
  JS, and DU is partially supported by the National Science Foundation
  under Grants No.\ 1615105, 1615109, and 1653300. FB acknowledges funding
  from the European Union's Horizon 2020 research and innovation programme
  (grant agreement No 726384). APSH is a fellow of the International Max
  Planck Research School for Astronomy and Cosmic Physics at the University
  of Heidelberg (IMPRS-HD). SCOG acknowledges support from the DFG via SFB
  881 ``The Milky Way System'' (sub-projects B1, B2 and B8).  JMDK
  gratefully acknowledges funding from the European Research Council (ERC)
  under the European Union's Horizon 2020 research and innovation programme
  via the ERC Starting Grant MUSTANG (grant agreement number 714907).  JMDK
  and MC gratefully acknowledge funding from the German Research Foundation
  (DFG) in the form of an Emmy Noether Research Group (grant number
  KR4801/1-1). SEM acknowledges funding during part of this work from the
  Deutsche Forschungsgemeinschaft (DFG) via grant SCHI 536/7-2 as part of
  the priority program SPP 1573 "ISM-SPP: Physics of the Interstellar
  Medium."
\end{acknowledgements}

\bibliographystyle{aa} %
\bibliography{m74brightestcloud} %

\appendix{}

\section{Total power data reduction}
\label{app:tp}

We\footnote{Cinthya Herrera wrote the scripts guided by Jérôme Pety. Chris
  Faesi and Antonio Usero extensively use these scripts and proposed
  important upgrades.  } have developed a set of customized scripts to
carry out ALMA single dish (``total power'' or TP) data reduction. These
scripts play a crucial role in the PHANGS-ALMA reduction pipeline as the
short-spacings often contain a significant fraction of the flux of nearby
galaxies. They are based on the ALMA calibration scripts, with
modifications informed by our experience carrying out and supporting single
dish data reduction using the GILDAS software for IRAM 30-m data. The main
aim of the scripts is to allow the user to define a fixed velocity window
for the baseline subtraction while also producing a suite of diagnostic
plots that allow assessment of imaging quality. This Appendix describes the
structure and usage of these scripts.

The scripts have been provided to the ALMA Regional Centers and the Joint
ALMA Observatory, where they have helped inform development of the
Observatory data reduction scheme. They are available on GitHub at
\url{https://sites.google.com/view/phangs/home/data}.

They were developed to be used in CASA version 4.7.2 that used the ASAP
(ATNF Spectral Analysis Package)
extension to deal with Total Power gridding and baselining. Since
then, newer versions of CASA contain dedicated tasks that performs these
tasks. We did not port our scripts yet.

The script distribution contains three files.
\begin{description}
\item[\bf \texttt{analysis\_scripts}] ALMA-provided folder that contains
  useful tools for TP ALMA data reduction. The scripts in this folder
  should not be modified.
\item[\bf \texttt{ALMA-TP-tools.py}] Script which contains our generic
  procedures for total power data reduction and imaging. This script should
  not be modified.
\item[\bf \texttt{galaxy-specific-scripts}] Folder containing the
    individual galaxy-specific scripts, \verb|GalaxyName-input.py|, which
  are a wrapper for the two previous scripts. Here the user specifies
  various galaxy- and project-specific imaging parameters, including phase
  center position, source systemic velocity, rest frequency, channel width,
  final cube velocity width, and the definition of baseline subtraction
  windows (see additional details in the end-to-end reduction description
  below).
\end{description}
The distributed code has to be saved in a folder named \verb|scripts_TP|
folder. The galaxy-specific \verb|GalaxyName-input.py| CASA script should
be executed in this folder. Two additional folders will be created at the
same level as the \verb|scripts_TP| one, \verb|data| and \verb|tmp|. The
first will contain the output cube fits files. The second is a temporary
folder where the TP data reduction happens and intermediate files are
stored. Once the data reduction is done, this folder can be deleted. Total
Power raw data stored in the original ALMA directory structure are first
replica in the \verb|tmp| folder. Only the needed files are copied. Data
reduction products are created in the \verb|calibration|
folder. Specifically, two additional subfolders are created, \verb|plots|
and \verb|obs_lists|. The \verb|plots| folder contains all plots created by
the data reduction scripts, for example the $T_{\rm sys}$ and the baseline
correction plots. The quality of the data can be judged checking these
plots. The \verb|obs_lists| folder contains the observation lists for the
data.

The main script, \verb|ALMA-TP-tools.py|, contains the procedures and tasks
for the data reduction. We now describe the main steps in order of their
execution during the processing.
\begin{description}
\item[\bf Import data to MS and split by antenna] The raw ALMA data are
  stored in the Archival Science Data Model (ASDM) format. The data are
  converted to measurement sets (MS), a CASA table format, using the task
  \verb|importasdm|.  Initial flagging is applied with the task
  \verb|flagcmd|. These are {\it a priori} flags, which are already encoded
  in and read from the MSs themselves. These flags could include issues
  such as when the mount is off source, calibration device is in an
  incorrect position, or power levels are not optimized. We also flag 7m
  antennas that may have been used for Total Power observations. Using the
  task \verb|sdsave|, each execution block data set in the MS format is
  split by antenna and converted to the ASAP format.
\item[\bf Generate $T_{\rm sys}$ calibration tables and apply additional
  flags] For each ASAP file, a system temperature ($T_{\rm sys}$)
  calibration table is generated using the task \verb|gencal|, which is
  needed to apply the calibration scale to the observations. One
  $T_{\rm sys}$ solution is created per spectral window.  In this step,
  additional flagging to the data can also be applied by creating a script
  containing custom flag commands. This script should be placed in the
  \verb|galaxy-specific-scripts/flags-folder| folder.
\item[\bf Subtract the atmospheric total power and convert to Kelvin] The
  $T_{\rm sys}$ calibration solution is applied to the data with the task
  \verb|sdcal2|. At the same time, this task subtract the contribution of
  the atmosphere to the total power (subtraction of the OFF measurement)
  and it corrects for the chromatic response of the atmosphere and the
  receivers by dividing by the OFF measurement. The output will thus be
  expressed in units of antenna temperature as
  \begin{equation}
    T_{\rm a}^{\star}=T_{\rm sys} \times \frac{(\emr{ON}-\emr{OFF})}{\emr{OFF}}.
  \end{equation}
\item[\bf Extract the frequency window that contains the emission line]
  From this step on-wards, we choose to continue the data processing only
  for the spectral window that contains the emission line of interest plus
  a baseline margin on each side of the line. The user must define in the
  script \verb|GalaxyName-input.py| both the rest frequency of the emission
  line and the velocity range of the cube that will be extracted from the
  data. The systemic velocity of the source is read from the MS. As the
  ALMA data are originally observed in the TOPO frame, the spectral axis is
  converted to the LSRK velocity frame at this stage to simplify the
  extraction. This uses functionality inside the \texttt{analysis\_scripts}
  folder. The extraction of the cube is done with the task \verb|sdsave|.
\item[\bf Baseline the spectra] We fit a polynomial though the spectrum
  parts that belong to the velocity ranges on both side of the lines. The
  velocity range to be ignored in this baseline fitting is defined in the
  \verb|GalaxyName-input.py| script. More than one velocity window can be
  defined. The order of the baseline fit is one (i.e., a linear fit) by
  default, but this can also be changed in the \verb|GalaxyName-input.py|
  script. Baseline fitting and subtraction call the the \verb|sdbaseline|
  task.
\item[\bf Convert to Jy/Beam and concatenate data from all antennas] After
  all the above steps for data calibration are done, the ASAP files for all
  execution blocks and antenna are converted back to MS format using the
  task \verb|sdsave|. For each execution block and individual antenna, the
  data are converted from Kelvin to Jansky using the factor measured and
  delivered by ALMA based on independent observations of point sources of
  known flux. This makes an higher order correction to the flux scale. All
  the data are then concatenated using the task \verb|concat|.
\item[\bf Grid the spectra onto a pixelized cube] All the execution blocks
  together are imaged together. The user can define the imaging parameters
  such as the projection center, the velocity frame of refence, the total
  velocity width, and the channel width in the \verb|GalaxyName-input.py|
  script. The gridding is done using the \verb|sdimaging| task,
  in the LSRK velocity reference frame using a spheroidal function.
\item[\bf Export the position-position-velocity to a fits file] The task
  \verb|exportfits| is used to export the signal and weight cubes to fits
  files.
\end{description}
In a few cases, an ozone telluric line were not satisfactorily subtracted
during the ON-OFF processing because the atmosphere contribution to the
total power varied too quickly. We used advanced baselining functionality
in the \texttt{GILDAS/CLASS} software to remove this artifact and to obtain
flat baselines. A deeper analysis that will be presented elsewhere (Usero
et al., in prep) showed that a change of observing mode is sometimes
required to get rid or at least decrease the contribution from some ozone
lines.

\section{Headlight spatial distributions of multiple tracers}

\FigMultiLines{} %

Figure~\ref{fig:cloudtracerstpeak} shows the spatial distribution at full
angular resolution of the peak brightness temperature around the headlight
for the observed molecular lines: CO(2$-$1), CO(1$-$0), $^{13}$CO(1$-$0),
CS(5$-$4), HCO$^+$(1$-$0), HCN(1$-$0), and
HNC(1$-$0). Figure~\ref{fig:cloudtracersIconv} shows the spatial
distribution of the emission integrated over the line profiles for the same
tracers, after convolution to the worst angular resolution obtained, i.e.,
the $^{13}$CO(1$-$0). These figures shows that HCO$^+$(1$-$0), HCN(1$-$0),
and HNC(1$-$0) are clearly detected in contrast with CS(5$-$4). The high
density tracers are slightly offset compared to the CO isotopologue
lines. The relatively low signal-to-noise of the HCO$^+$, HCN, and HNC
lines makes the origin of the offsets uncertain.

\end{document}